# $Sb_2Se_3$ and $SbBiSe_3$ Surface Capping and Biaxial Strain Co-Engineering for Tuning the Surface Electronic Properties of $Bi_2Se_3$ Nanosheet- A Density Functional Theory based Investigation

Naresh Bahadursha [a], Banasree Sadhukhan[b,c], Tanay Nag[d], Swastik Bhattacharya[d], and Sayan Kanungo*[a,e]

*Corresponding authors: E-mail: sayan.kanungo@hyderabad.bits-pilani.ac.in (SK)

[a.] Electrical and Electronics Engineering Department, Birla Institute of Technology and Science-Pilani, Hyderabad Campus, Hyderabad-500078, India.
[b.] Department of Physics and Nanotechnology, Faculty of Engineering & Technology, SRM Institute of Science and Technology, Kattankulathur, 603203, Chennai, Tamil Nadu, India
[c.] Tata Institute of Fundamental Research, Hyderabad, Telangana-500046, India
[d.] Physics Department, Birla Institute of Technology and Science-Pilani, Hyderabad Campus, Hyderabad-500078, India
[e.] Materials Centre`for Sustainable Energy & Environment, Birla Institute of Technology and Science-Pilani, Hyderabad Campus

**Abstract**

In this work, for the first time, a density functional theory (DFT) based comprehensive theoretical study is performed on the surface electronic properties of $Bi_2Se_3$ nanosheet in the presence of a surface capping layer as well as mechanical strain. The study systematically introduces a biaxial compressive and tensile strain up to 5% in natural, $Sb_2Se_3$ surface capped, and $SbBiSe_3$ surface capped $Bi_2Se_3$, and the subsequent effects on the electronic properties are assessed from the surface energy band (E-k) structure, the density of states (DOS), band edge energy and bandgap variations, surface conducting state localization, and Fermi surface spin-textures. The key findings of this work are systematically analyzed from conducting surface state hybridization through bulk in the presence of surface capping layers and applied biaxial strain. The result demonstrates that the interplay of surface capping and strain can simultaneously tune the surface electronic structure, spin-momentum locking results from change in electronic localization and interactions. In essence, this work presents an extensive theoretical and design-level insight into the surface capping and biaxial strain co-engineering in $Bi_2Se_3$, which can potentially facilitate different topological transport for modern optoelectronics, spintronics, valleytronics, bulk photovoltaics applications of engineered nanostructured topological materials in the future.

## 1. Introduction

The discovery and successful realization of topological insulator (TI) is considered one of the recent notable breakthroughs in condensed matter physics. The TI exhibits a non-zero bulk bandgap, wherein conductive surface states exists within the bulk bandgap energy range [1]. These symmetry dependent and topologically protected surface states demonstrate a number of exotic properties, such as linear E-k relationship near the Dirac point, spin-momentum locked surface states, dissipation-less surface transport [1-3]. Such distinct properties of TIs are considered highly potential for a number of conventional and emerging technological applications including electronics [1], optoelectronic [4-5], photovoltaic [6-7], magnetoelectronic [8], spintronic [8-9], and topological transport applications [10-12]. Consequently, to develop a detailed understanding of any TI and its potential applications, it is essential to study the electronic structures and predict how the same can be efficiently tuned using different material engineering strategies.

One prominent example of three-dimensional (3D) TIs are some specific members of the tetradymite group of materials that are represented as $M_2X_3$, wherein $Bi_2Se_3$, $Bi_2Te_3$, and $Sb_2Te_3$ are identified as TI [13]. The crystal structure of these $M_2X_3$ TIs exhibits a layered structure where each unit layer consists of five atomic layers of covalently bonded M (Bi or Sb) and X (Se or Te) atoms in the X(1)–M(2)–X(3)–M(4)–X(5) sequence known as quintuple layers (QL) [14]. In bulk $M_2X_3$ TIs, individual QLs are held together by the vdW force. The Bismuth Selenide ($Bi_2Se_3$) is considered as one of the ideal 3D TIs, with topologically protected single surface Dirac cone and bulk direct bandgap around 0.3 eV [15-16]. Specifically, the presence of a single Dirac cone in pristine $Bi_2Se_3$ has drawn further attention for realizing electronic devices for quantum computation [17] as well as other topological devices exploiting dissipation-less in-plane carrier transport [15]. These factors have recently driven an increasing research interest in $Bi_2Se_3$ using both experimental characterization and ab-initio theoretical techniques. Specifically, tuning the surface electronic properties of $Bi_2Se_3$ using different material engineering strategies, including surface modification, alloy formation, doping, and mechanical strain appears at the forefront of such research.

The $Bi_2Se_3$ is experimentally synthesized using different synthesis routes, including molecular beam epitaxy [18-20], vapor phase deposition [15], chemical vapor deposition [21-22], hybrid physical chemical vapor deposition [23], where the nanosheet formation is one of the most common structural form of grown $Bi_2Se_3$ with few QL thickness [18-20, 22, 24]. The presence of a topological single surface Dirac cone in $Bi_2Se_3$ was experimentally demonstrated from the angle-resolved photoemission spectroscopy and Shubnikov-de Haas (SdH) oscillations [15, 25]. Interestingly, it has been observed that the as-synthesized $Bi_2Se_3$ demonstrates Selenium (Se) vacancy-mediated n-type nature, preserving the surface Dirac cone [15]. In contrast, a p-type nature can be induced by small substitutional Calcium (Ca) doping in $Bi_2Se_3$, where the presence of surface Dirac cone near Fermi-level strongly depends on doping concentrations [26]. Moreover, it has been conclusively established that the presence of a trace quantity of substitutional Ca doping and surface hole doping with $NO_2$ preserves the helical nodal Dirac fermion on the surface of $Bi_2Se_3$ with tunable topological fermion density in the vicinity of the Kramer's point, and allow spin-polarized topologically protected edge channels at room temperature [2]. On the other hand, it is observed that the presence of intrinsic periodic mechanical strain due to edge dislocations at the surface of $Bi_2Se_3$ can notably affect the surface Dirac states and modify the band topology, wherein a small bandgap opening is observed for surface tensile strain (compressive stress) and energy of the gapless Dirac state shifts with surface compressive strain (tensile stress) [18]. Furthermore, reversible tuning of surface Dirac state energy of $Bi_2Se_3$ are demonstrated using externally applied compressive strain (tensile stress) in the lattice [24].

The aforementioned experimental demonstrations of surface electronic structure tuning in $Bi_2Se_3$ using substitutional doping, surface functionalization, and mechanical strain suggest the need to develop detailed theoretical insight into the surface electronic properties of $Bi_2Se_3$ under the influence of different material engineering techniques. In this context, the density functional theory (DFT) based ab initio calculations are becoming instrumental for the theoretical characterization of the electronic structures of engineered TIs [27-29], and is often considered to complement the analysis of experimental findings on TIs [18,24,30-33]. Driven by this paradigm, recently, a number of reports have attempted to theoretically investigate the surface electronic properties of doped, surface modified, and strained $Bi_2Se_3$. Specifically, the topological surface state tuning using an external electric field in Cr-doped $Bi_2Se_3$ [34], ionic bond induction through Sn-doping [35], time-reversal symmetry (TRS) breaking with magnetic transition metal (Mn, Fe, Co, Ni) doping [36], as well as magnetic non-metallic N-doping [37] was previously reported. Alternatively, the introduction of a surface capping layer on $Bi_2Se_3$ has shown notable promise for tuning the surface electronic properties of $Bi_2Se_3$ [38-39]. Specifically, the introduction of magnetic capping layers like $CrI_3$ [38], and EuS [39] can break the TRS in $Bi_2Se_3$ leading to significant modulation in its surface states, without the possibility of inadvertently compromising the topological features as in the case of substitutional doping. On the other hand, the previously reported DFT studies reveal that the surface Dirac states of $Bi_2Se_3$ are highly sensitive towards the applied mechanical strain, where the extent of surface state electronic structure modulation depends on the nature (uniaxial/biaxial or tensile/compressive) and magnitude of the applied strain in the lattice [40-42]. Specifically, it has been observed that a relatively larger uniaxial compressive strain (tensile stress) across the out-of-plane direction can potentially annihilate the Dirac point and introduce a finite bandgap of ~5 meV [40,43]. On the other hand, the application of strain can also significantly influence the spin-orbital interaction in $Bi_2Se_3$, and can modify the surface spin texture [42]. However, to the best of author's knowledge, to date, there are no theoretical or experimental reports available on the simultaneous effects of surface capping and strain engineering on $Bi_2Se_3$. Since, both surface capping layer engineering and strain engineering exhibit significant influences on the surface electronic properties, it can be surmised that the simultaneous interplay between both of these engineering strategies can potentially lead to emergent surface electronic structures of $Bi_2Se_3$.

Driven by this paradigm, in this work, for the first time, the effects of surface capping layer and strain co-engineering is theoretically studied on the surface electronic properties of $Bi_2Se_3$ nanosheet using ab-initio DFT calculation. For this study, the non-topological $Sb_2Se_3$ surface capping layer belonging to tetradymite group with hexagonal lattice structure and comparable lattice constants is considered. Moreover, an alloy of $Sb_2Se_3$ and $Bi_2Se_3$, i.e. $SbBiSe_3$ having one Bi and one Sb atom in the surface QL, is also considered as a surface capping layer to study the intermediate effects of surface layer engineering $Bi_2Se_3$. Consequently, the work also represents the first attempt to study the influence of non-magnetic and non-topological surface capping layers on the surface electronic properties of $Bi_2Se_3$. In this effect, the effects of different biaxial compressive and tensile strain are extensively studied on the surface electronic properties of natural and surface capped $Bi_2Se_3$, and the key findings are systematically corelated with the atomic orbital interactions. Specifically, the work demonstrates that the introduction of surface capping layer and applied biaxial tensile strain (compressive stress) notably delocalizes the surface conducting state resulting in their hybridization though the bulk region of $Bi_2Se_3$ nanosheet, wherein the delocalized surface conducting states of surface-capped $Bi_2Se_3$ nanosheet can be re-localized by applying biaxial compressive strain (tensile stress). Subsequently, the surface capping/strain induced surface state hybridization leading to Dirac point annihilation, whereas Dirac point restoration can also be achieved by eliminating surface state hybridization in $Bi_2Se_3$ nanosheet. Apart from influencing the Dirac point and band edge energies, the surface capping/strain can significantly influence the spin-momentum locking near the surface band edges, and the interplay between these factors can lead to emerging spin chirality features in $Bi_2Se_3$ nanosheet.

Next, the organization of the following sections of this work is duly summarized here. In the following section, the computational methodologies, including material specifications, DFT simulation set-up, relevant parameter definitions, are discussed in detail, and the calculated results are duly calibrated with reported results to verify the reliability of the simulation framework.

The following section presents the findings of this work in terms of structural, surface electronic, and spin texture of surface capped and strain co-engineered $Bi_2Se_3$ material systems that are methodically analyzed from the atomistic properties of the materials. Finally, the key findings are summarized, and their broader significances are duly highlighted in the conclusion section.

## 2. Computational Details

In this section, the computational model of the material systems, the DFT calculation framework, and theoretical parameter definitions are discussed in detail.

### 2.1. Material Specifications

The Bismuth Selenide ($Bi_2Se_3$) is a layered material with individual layers consisting of five atoms in a Se-Bi-Se-Bi-Se arrangement that forms a quintuple layer (QL), where each QL layers are held together by relatively weaker vdW force [18]. The $Bi_2Se_3$ demonstrates a rhombohedral crystal structure and belongs to $R\vec{3}m$ space group [41,44]. The hexagonal symmetry in bulk $Bi_2Se_3$ lattice results in a hexagonal Brillouin zone (BZ). Correspondingly, in this work, the energy band structure is considered along the K-G-M high symmetry route of the BZ. The lattice structure and calculated band structure of bulk $Bi_2Se_3$ are depicted as supplementary information (**Fig. S1**). It should be noted that for bulk $Bi_2Se_3$, the calculated in-plane lattice constant value of 4.27 Å is slightly higher than the experimentally reported lattice constant value of 4.14 Å [36,45]. Moreover, the estimated indirect energy bandgap value of 0.19 eV for bulk $Bi_2Se_3$ is comparable with some previous theoretical calculation [45-46], but slightly lesser compared to the experimentally reported energy bandgap value between 0.22 to 0.30 eV [47-48].

For studying the surface properties of $Bi_2Se_3$ nanosheets, the slab model of 6 QL (with 30 atoms) in unit-cell (1×1×1) configuration is considered where top and bottom surfaces are constructed using the slab model in (0001) direction [40-41], as shown in **Fig.1**. In the slab model, a vacuum of 78 Å in the out-of-plane directions of the unit cell is considered to eliminate interaction of the periodic images along top and bottom surfaces. Owing to the negligible finite size effect, the 6 QL slab model configuration of $Bi_2Se_3$ offers an effectively zero-bandgap surface Dirac cone with a minimum number of atoms in the unit-cell [49], which can be verified from the surface electronic property calculation of 3QL, 4QL, 5QL, 6QL slab models of $Bi_2Se_3$ nanosheets depicted as supplementary information (**Fig. S2**), and are also consistent with previous DFT calculations [40-41].

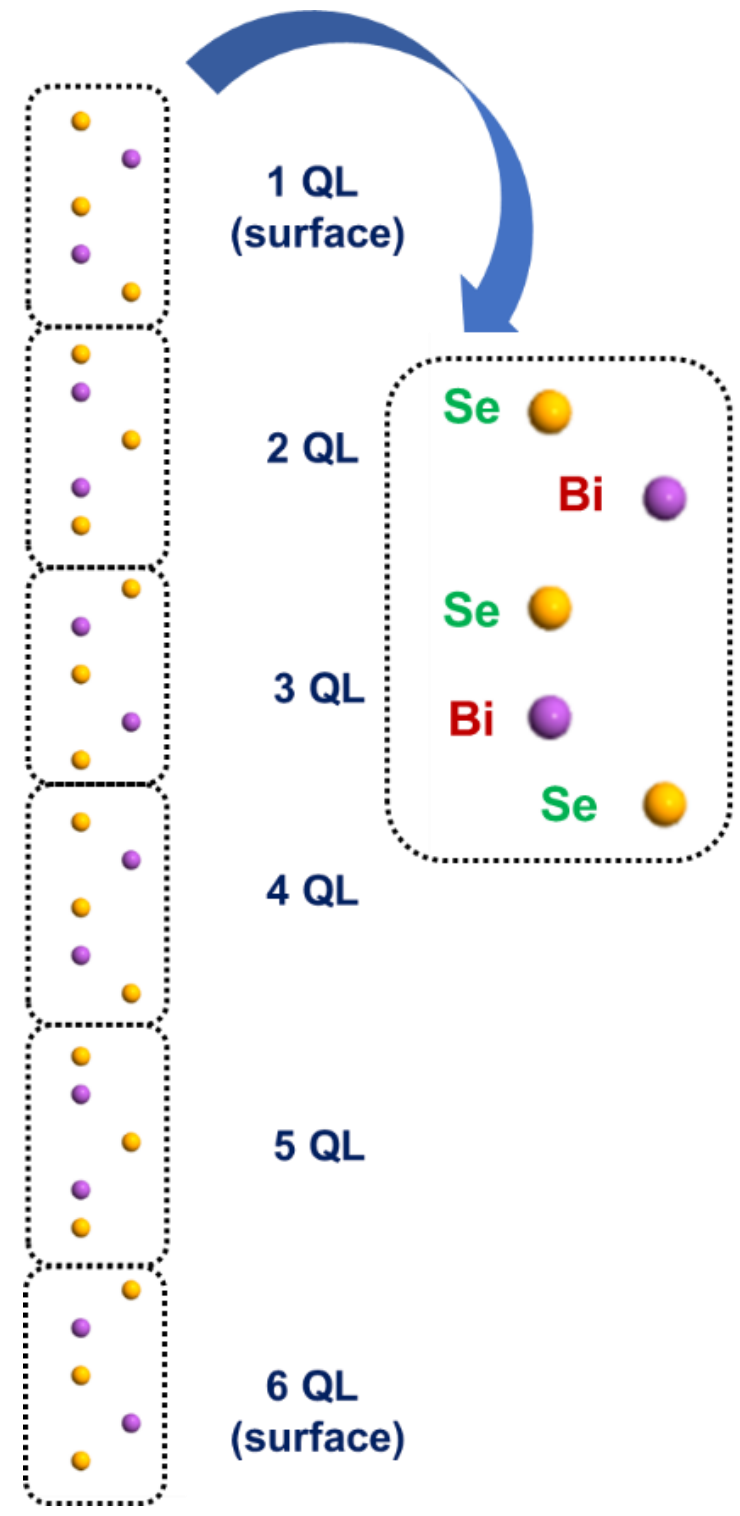

**Fig. 1.** The schematic representation of (a) 6QL of $Bi_2Se_3$ unit cell (1 ×1 ×1) from side-view.

It should be noted that the typical MBE-grown $Bi_2Se_3$ nanosheet thickness typically vary from 6QL - 10QL [19-20, 24], which is also consistent of the theoretical slab model of $Bi_2Se_3$ nanosheet considered in this work.  On the other hand, in this work, for the (0001) slab configuration of $Bi_2Se_3$ nanosheet, the DFT calculated Dirac point is energetically located slightly below the Fermi level. In this context, it is worth mentioning that the position of Dirac point in energy depends on the $Bi_2Se_3$ surface specification [43,50-51], specifically for (0001) surface, the Dirac point is located below Fermi-level [40,43,51-54]. Next, for this study, two artificial material systems are constructed by replacing the top and bottom QL of $Bi_2Se_3$ with $Sb_2Se_3$ and $SbBiSe_3$ in natural stacking orientation of $Bi_2Se_3$, resulting in $Sb_2Se_3$ and $SbBiSe_3$ surface capped $Bi_2Se_3$ material systems. The surface-capped material systems also have a 6QL configuration, where the specifications and atomic arrangement of top and bottom surface QLs are kept identical. Moreover, these artificial material systems are considered to retain the hexagonal lattice structure, and their energy band structure is also considered along the K-G-M high symmetry route of the BZ.

## 2.2. Simulation Setup

In this work, the atomistix tool kit (ATK) and virtual nano lab (VNL) graphical user interface

(GUI) from Synopsys Quantum wise are considered for performing the density functional theory (DFT) based first principle calculations [55]. The DFT calculations are performed in a linear combination of atomic orbital (LCAO) double-zeta polarized basis sets, with density mesh cut-off energy of 125 Hartree and a Monkhorst Pack grid of 12x12x1 for BZ sampling [55]. Next, to realize equilibrium configuration corresponding to the minimum force per atom and lowest ground-state energy, individual unit-cells are relaxed through geometry optimization considering the Limited memory Broyden Fletcher Goldfarb Shanno (LBFGS) algorithm with pressure and force tolerance of 0.0001 eV/Å$^3$ and 0.01 eV/Å, respectively [55]. For geometry optimization and electronic property calculations, the Generalized Gradient Approximation (GGA) DFT method with Purdew Burke and Ernzerhof (PBE) exchange-correlational functional are considered [55], which offers an efficient trade-off between computational efficiency and reliability for estimating the structural and surface electronic properties of topological insulators like $Bi_2Se_3$ [34,40]. Moreover, for electronic property calculation, the spin-orbit coupling (SOC) effects is enabled with GGA-PBE to incorporate the strong SOC effects at the surface band edges [40].

For the strain analysis, in this work, a biaxial strain of -5% to +5% is considered for every material configuration, where the strain (S) is calculated as:

$$S = 100 \times \frac{a_{strained} - a_{relaxed}}{a_{relaxed}} \% \tag{1}$$

In the **Eq. (1)**, $a_{strained}$ and $a_{relaxed}$ represent the in-plane lattice vectors of relaxed and biaxially strained configuration for any material system. The **s < 0**, represents an applied biaxial compressive stress, which leads to a biaxial tensile (BT) strain generation in the lattice [56]. In contrast, **s > 0**, indicates the presence of biaxial tensile stress and, thereby, biaxial compressive (BC) strain in the lattice. The applied strain range considered in this work is in agrees with the theoretical literature on $Bi_2Se_3$ [18,40].

The structural stabilities of the individual relaxed material systems are assessed in terms of the cohesive energy per unit atom ($E_{coh}$), which is defined as [57]:

$$\mathrm{E_{coh}} = \frac{\mathrm{E_{6QL}} - (\mathrm{n_1.E_{Bi}} + \mathrm{n_2.E_{Sb}} + \mathrm{n_3.E_{Se}})}{30} \tag{2}$$

In the **Eq. (2)**, $E_{6QL}$, $E_{Bi}$, $E_{Sb}$, and $E_{Se}$ represent the ground state energy (<0) of the 6QL of the material system, isolated Bismuth atom, isolated Antimony atom, and isolated Selenium atom, respectively. Moreover, $E_{coh} < 0$ represents a stable lattice formation, where a higher magnitude of $E_{coh}$ indicates better stability [58].

Finally, the valence charge at any specific atomic sites of any material system is calculated using the Mulliken charge analysis [59], and are corelated with the spatial distribution of Electron Localization Function (ELF). The ELF presents an efficient means for visualizing the degree of electron localization and therefore, the spatial electron distribution around any atomic site [60]. The ELF value can be between 0 to 1, wherein a higher value at any spatial location suggests a higher electron localization at that point [61].

## 3. Results and Discussions

In this section, the structural properties and surface electronic properties of the material systems are discussed in details, and in this context the effects of surface capping layers and biaxial strain are methodically analysed.

### 3.1. Structural Properties of Natural and Surface Capped $Bi_2Se_3$ Nanosheets

To analyse the effects of surface capping layers, the natural, $Sb_2Se_3$ surface capped ($Sb_2Se_3/Bi_2Se_3$), and $SbBiSe_3$ surface capped ($SbBiSe_3/Bi_2Se_3$) $Bi_2Se_3$ nanosheets are considered, as depicted in **Fig.2**. Moreover, the ELF profile of the top-surface layer with unit cell lattice vectors, Bi (Sb)-Se bond lengths, and cohesive energy values are also incorporated for individual material systems in **Fig.2**. It should be noted that for $SbBiSe_3$ capping layer, two distinct interlayer stacking configurations can be identified based on the position of Sb atom in the surface QL of $SbBiSe_3$. Specifically, in QL of $SbBiSe_3$ QL of $SbBiSe_3$, if the Sb atom is away from the surface and near the surface, the subsequent configurations are termed configuration-1 and configuration-2, as shown in **Fig.2(c) and (d)**, respectively.

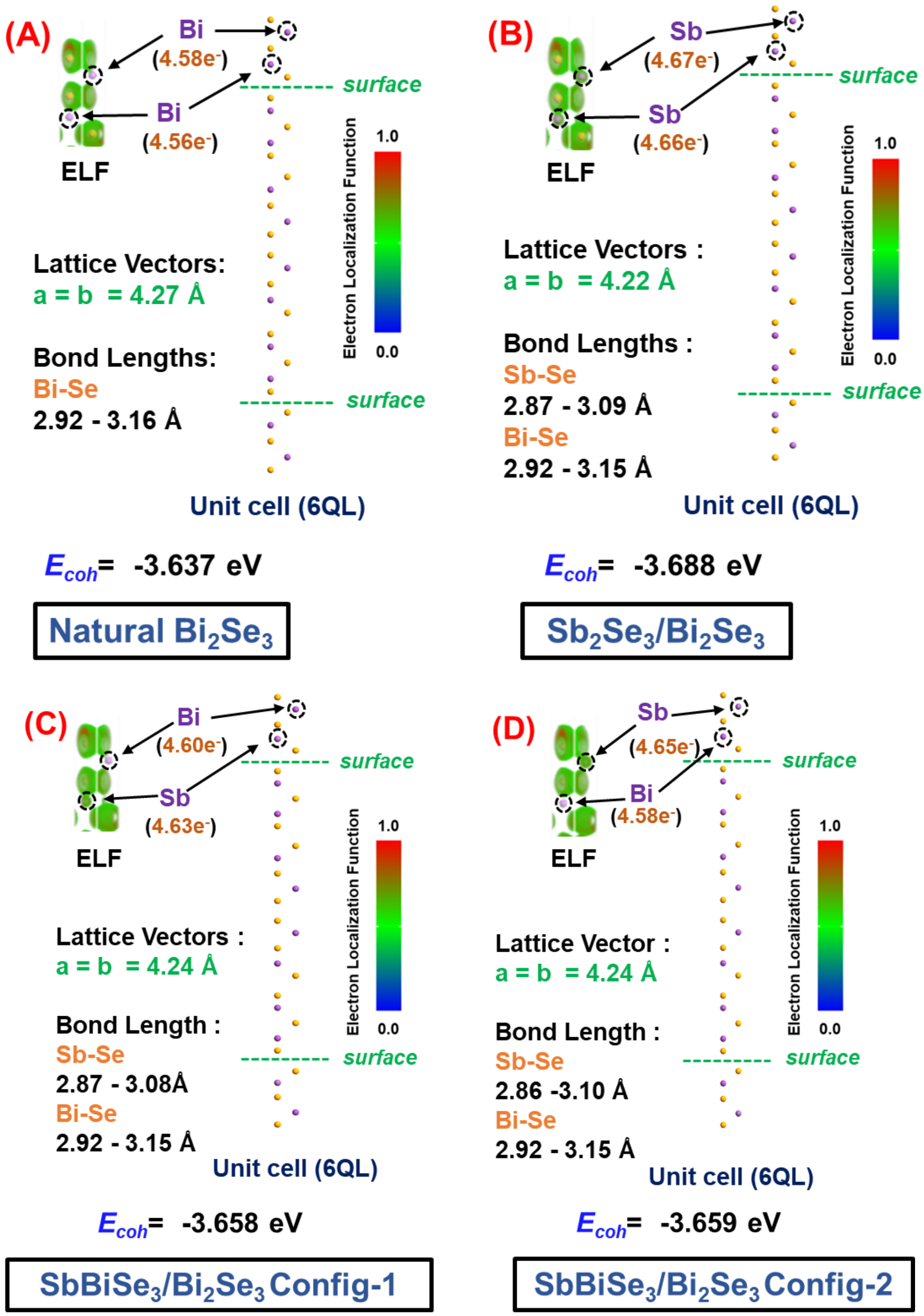

**Fig. 2.** Schematic representation, structural parameters, and cohesive energy of (a) natural $Bi_2Se_3$, (b) $Sb_2Se_3$-capped $Bi_2Se_3$, (c) $SbBiSe_3$-capped $Bi_2Se_3$ in configuration-1 and (d) $SbBiSe_3$-capped $Bi_2Se_3$ in configuration-2 from side-view.

The cohesive energy values in **Fig.2** suggest that all the material systems are sufficiently stable in nature ($|E_{coh}| >> 0.026$ eV, i.e. KT @ T= 300K). Interestingly, the $|E_{coh}|$ slightly increases

with $Sb_2Se_3$ and $SbBiSe_3$ surface capping over $Bi_2Se_3$. Structurally, the relatively smaller atomic radius of Sb (1.33Å [62]) compared to Bi (1.43 Å [62]) results in a smaller Sb-Se bond lengths (2.86 – 3.10 Å), compared to Bi-Se bond length (2.92 – 3.16 Å) in the surface QLs. However, in the presence of an Sb atom in surface QLs of $Sb_2Se_3/Bi_2Se_3$ and $SbBiSe_3/Bi_2Se_3$, the Bi-Se bond lengths remain comparable to that of natural $Bi_2Se_3$. Consequently, the overall in-plane lattice constant of $Sb_2Se_3/Bi_2Se_3$ reduces from 2.27 Å to 2.22 Å, resulting in a small (~2%) biaxial compressive stress (tensile strain) in bulk $Bi_2Se_3$ QLs. Similarly, an intermediate in-plane lattice constant of 2.24 Å is observed for both $SbBiSe_3/Bi_2Se_3$ lattice configurations, corresponding to even smaller (~1%) biaxial compressive stress (tensile strain) in bulk $Bi_2Se_3$ QLs. On the other hand, the ELF spatial profile indicates a relatively larger electron localization around the Sb atom compared to the Bi atom in surface QLs of $Sb_2Se_3/Bi_2Se_3$ and $SbBiSe_3/Bi_2Se_3$, which is consistent with the larger Mulliken charge at Sb atom (4.66$e^-$ to 4.67$e^-$) compared to Bi atom (4.56$e^-$ to 4.60$e^-$). This can be attributed to the relatively higher electronegativity and smaller atomic radius of the Sb atom (Pauling electronegativity 2.05) compared to the Bi atom (Pauling electronegativity 2.02), allowing relatively larger electronic accumulation around the Sb atom. Interestingly, in surface $SbBiSe_3$ QL, the presence of Sb atom also slightly increases the Mulliken charge of Bi in that layer compared to a surface $Bi_2Se_3$ QL.

Therefore, the results suggest that the introduction of an Sb in the surface QL modifies the atomic orbital overlap and electronic distributions in the surface QLs, which is expected to induce appreciable changes in the surface electronic properties of $Sb_2Se_3/Bi_2Se_3$ and $SbBiSe_3/Bi_2Se_3$, compared to natural $Bi_2Se_3$.

### 3.2. Electronic Properties of Natural and Surface Capped $Bi_2Se_3$ Nanosheets

The surface electronic properties of natural and surface-capped $Bi_2Se_3$ nanosheets are analysed from the energy band (E-k) structure, the density of states (DOS), and atomic orbital projected band structures and are depicted in **Fig.3** to **Fig.6**. Moreover, topological insulator like $Bi_2Se_3$ is characterized by its highly localized conducting surface states with helical spin texture, where the spin of an electron is locked at a right angle with the momentum of conducting surface states, i.e. spin-momentum locking. Consequently, the effects of surface capping are also considered in this context. To this effect, the out-of-plane projected surface Bloch states (at 'G' point) across the 6QL slabs and Fermi-surfaces (at 'G' point) with their spin projections on in-plane momentum component are also depicted in **Fig.3** to **Fig.6**.

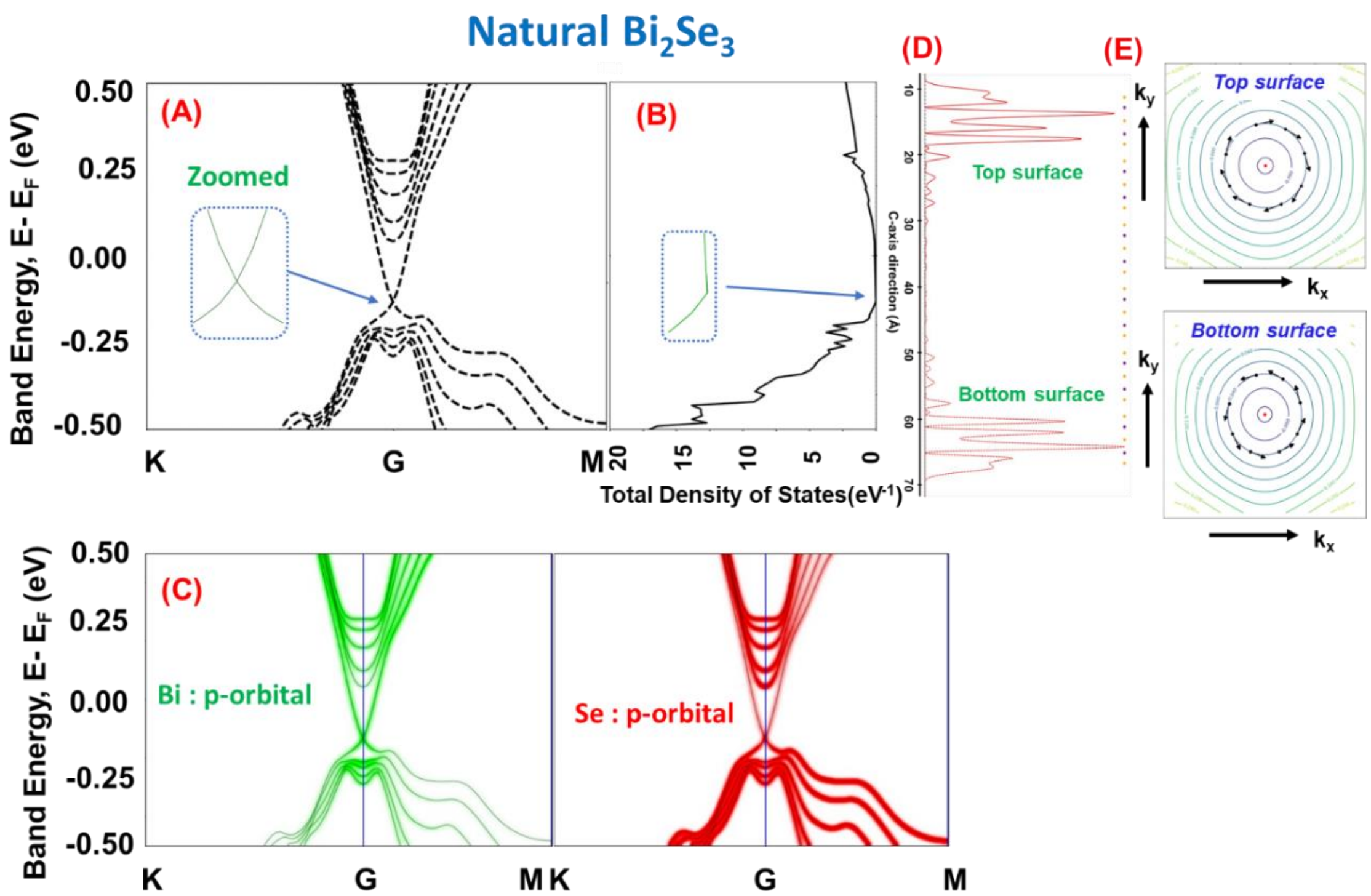

**Fig. 3.** Plots of (a) energy band structure, (b) density of states, (c) atomic orbital projected band structures, (d) surface Bloch states projections (along the C-axis), (e) Fermi surfaces on the Dirac cone (at G-point) of natural $Bi_2Se_3$.

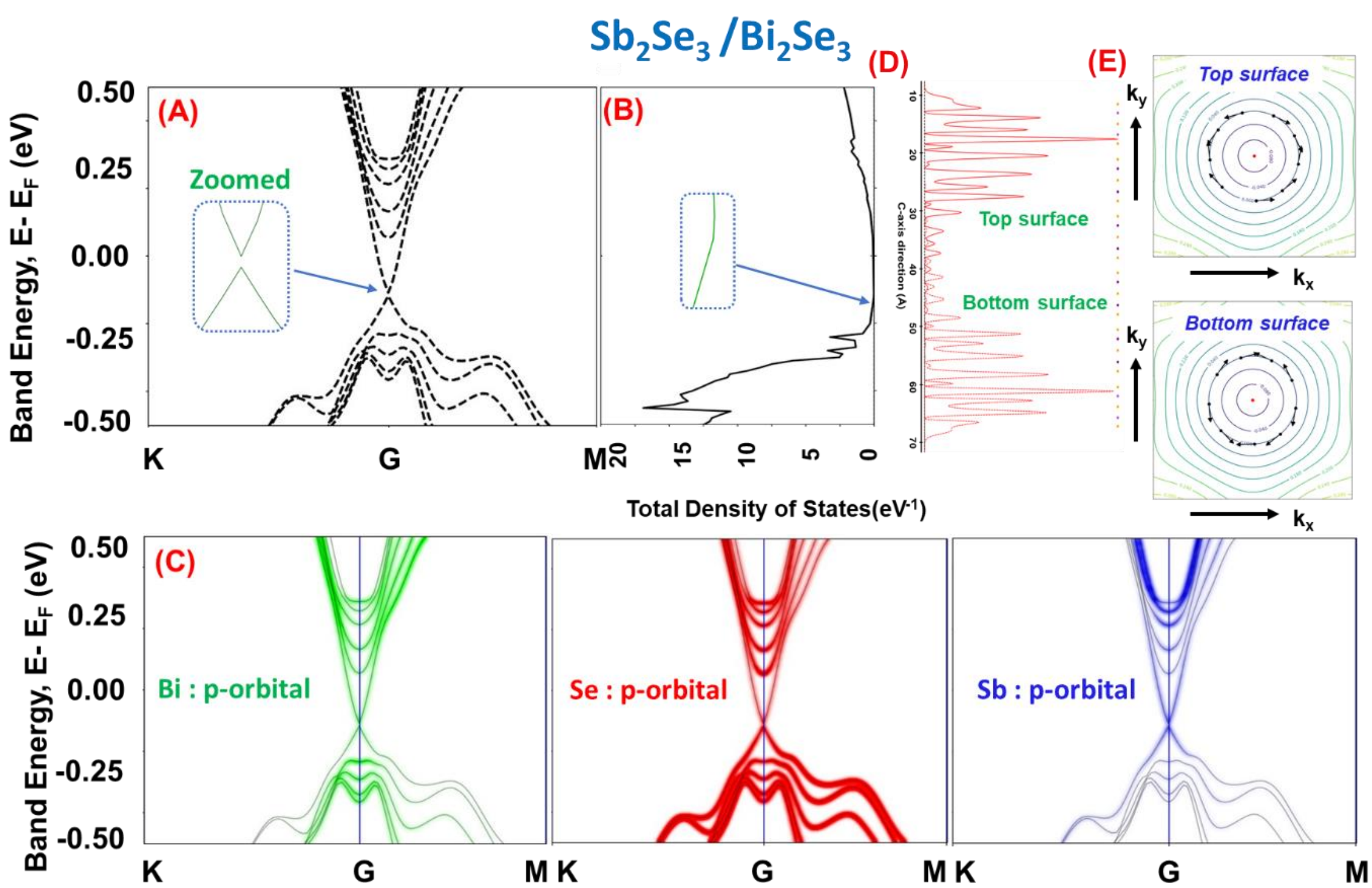

**Fig. 4.** Plots of (a) energy band structure, (b) density of states, (c) atomic orbital projected band structures, (d) surface Bloch states projections (along the C-axis), (e) Fermi surfaces on the band edge (at G-point) of $Sb_2Se_3/Bi_2Se_3$.

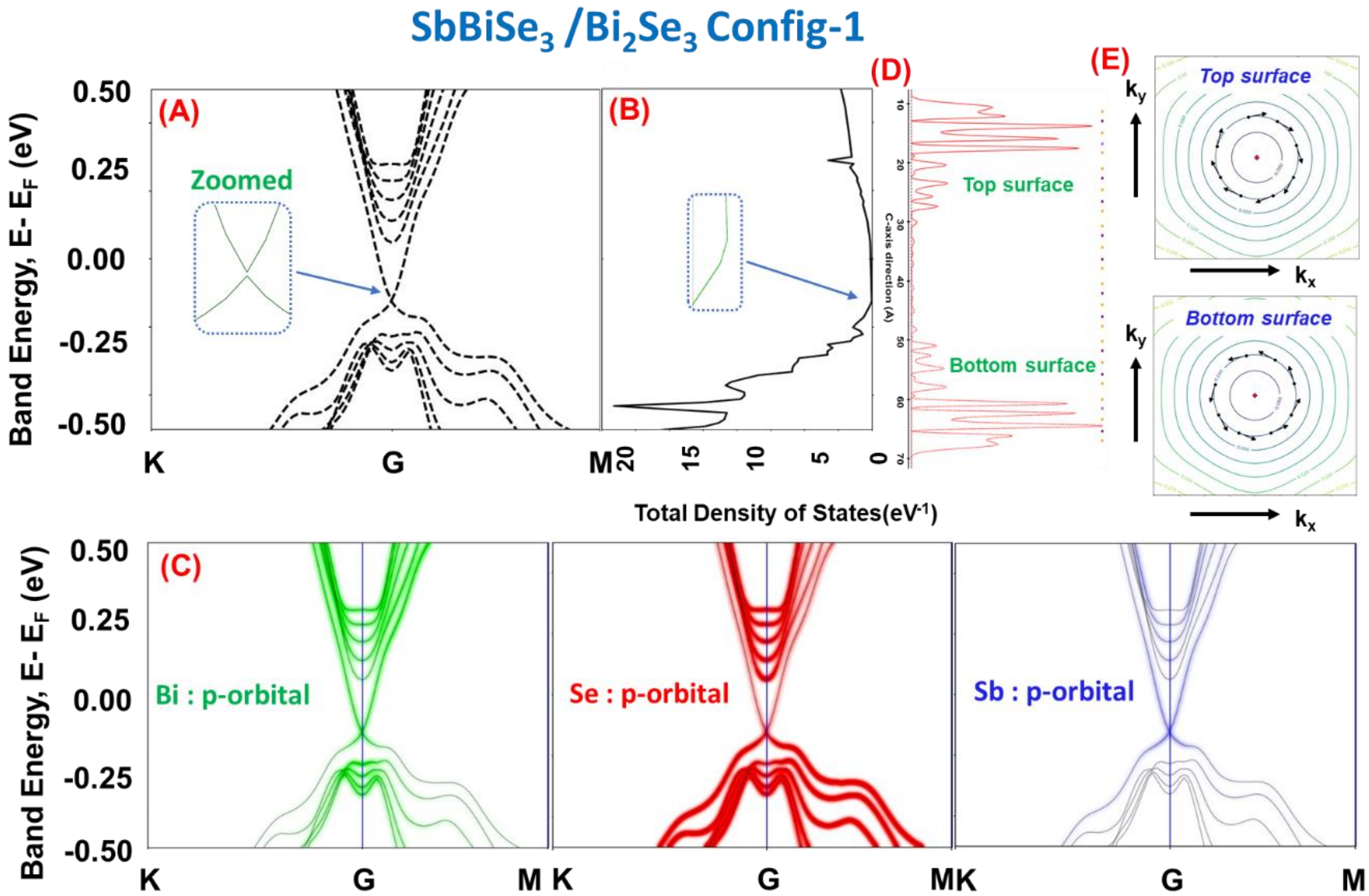

**Fig. 5.** Plots of (a) energy band structure, (b) density of states, (c) atomic orbital projected band structures, (d) surface Bloch states projections (along the C-axis), (e) Fermi surfaces on the band edge (at G-point) of $SbBiSe_3/Bi_2Se_3$ config-1.

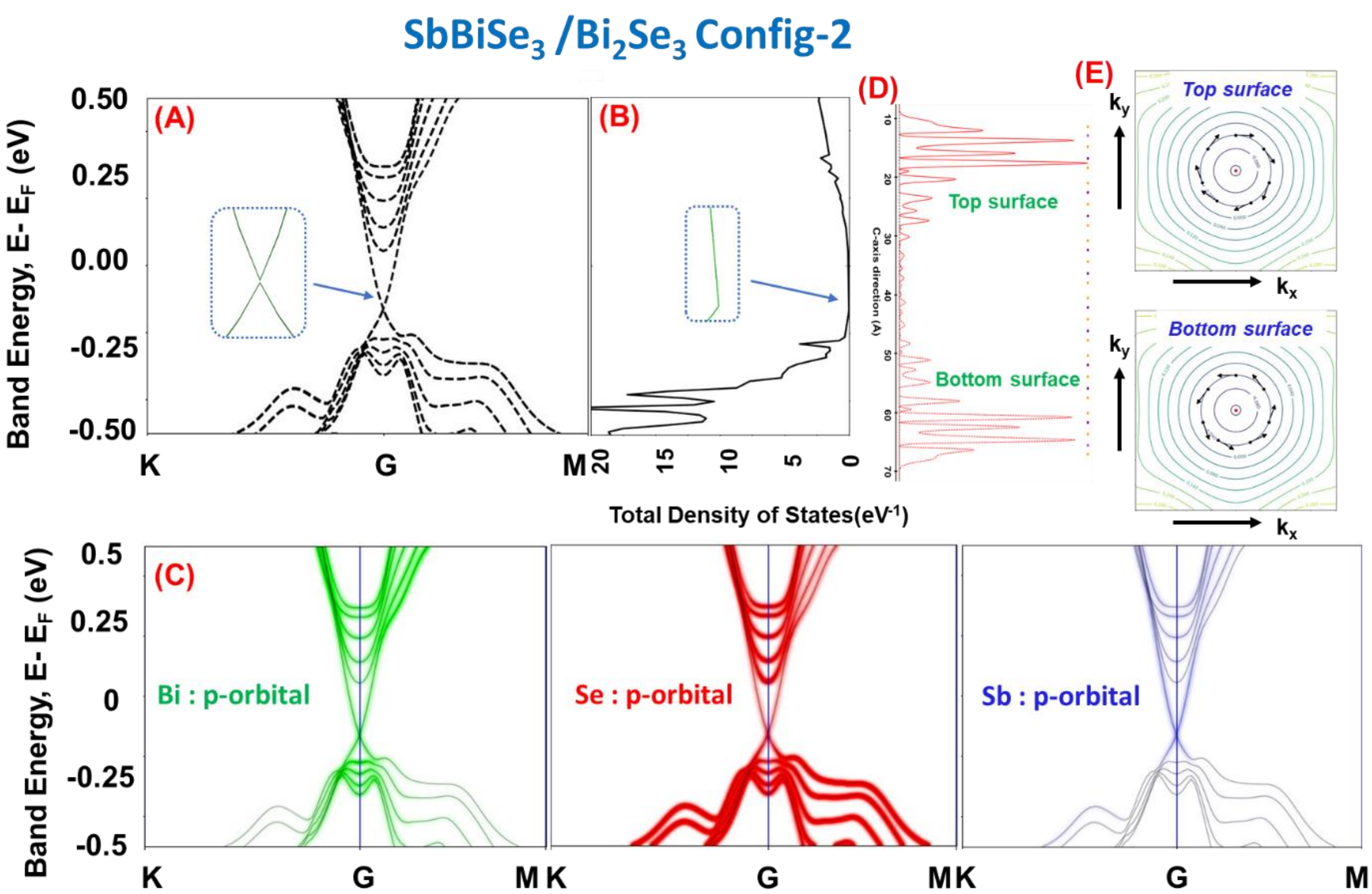

**Fig. 6.** Plots of (a) energy band structure, (b) density of states, (c) atomic orbital projected band structures, (d) surface

Bloch states projections (along the C-axis), (e) Fermi surfaces on the band edge (at G-point) of $SbBiSe_3/Bi_2Se_3$ config-2.

**Fig.3(a)** indicates that the natural $Bi_2Se_3$ nanosheet exhibits a gapless Dirac point at the 'G' high-symmetry point of the BZ with a linear E-k dispersion relation near the Dirac point, which is a typical characteristic of topological surface states. Interestingly, the introduction of $Sb_2Se_3$ and $SbBiSe_3$ surface capping layers retains the qualitative band structure of natural $Bi_2Se_3$ with linear E-k dispersions near the 'G' point, as depicted in **Fig.4(a)**, **Fig.5(a)**, and **Fig.6(a)**. However, these surface capping layers annihilate the Dirac point by inducing finite energy bandgaps, wherein conduction band maxima (CBM) and valence band minima (VBM) moves away from each other at the 'G' point. Specifically, the $Sb_2Se_3$ and $SbBiSe_3$ surface capping layers lead to a bandgap of ~ 9 meV and ~ 2 to 3 meV, respectively. Moreover, **Fig.3(b)**, **Fig.4(b)**, **Fig.5(b)**, and **Fig.6(b)** indicate that the introduction of surface capping layers tends to modify the E-k curvature of the bands below the VBM, leading to appreciable and distinct changes in the DOS profiles in the valence band of $Sb_2Se_3$/ $Bi_2Se_3$ and $SbBiSe_3$/ $Bi_2Se_3$.

The surface QL projected band structures of each material system are illustrated as supplementary information (**Fig. S3**). The results reveal that compared to natural $Bi_2Se_3$, notably smaller surface QL layer contributions can be observed near the 'G' point of $Sb_2Se_3$/ $Bi_2Se_3$, whereas, the surface QL contributions are slightly lesser for $SbBiSe_3$/ $Bi_2Se_3$. The atomic orbital projected band structure in **Fig.3(c)**, **Fig.4(c)**, **Fig.5(c)**, and **Fig.6(c)** suggests that the p-orbital of Bi and Se predominantly populates the energy states at the Dirac point of natural $Bi_2Se_3$ nanosheet, whereas a notable contribution of the p-orbital of Sb atom can also be observed at the band edges of surface-capped $Bi_2Se_3$ nanosheet. Specifically, for $Sb_2Se_3$/ $Bi_2Se_3$, a relatively smaller Bi p-orbital contribution and relatively stronger Sb p-orbital contribution at the band edges can be observed compared to other material systems. In contrast, Sb p-orbital contribution (Bi p-orbital contribution) remains relatively smaller (larger) for both the configurations of $SbBiSe_3/Bi_2Se_3$. In natural $Bi_2Se_3$, the conduction band above the Dirac point is predominantly populated by the p-orbital of Bi, and the p-orbital of Se primarily contributed to the valence band above the Dirac point, which is consistent with the previous theoretical report [40]. Interestingly, in each nanosheet, contribution of the Se p-orbital to valence band is spread across the 'K'-'G'-'M' path. In contrary to this, the contribution of the Bi and Sb p-orbital in the valence band is localized around the 'G' point, wherein the contribution of Sb p-orbital can only be observed at the top of the valence band near the VBM. However, a larger contribution of Sb p-orbital is observed in the conduction band at a higher energy from the CBM and near the G point of $Sb_2Se_3/Bi_2Se_3$, wherein the Sb p-orbital contribution remain much lesser in the entire conduction band of $SbBiSe_3/Bi_2Se_3$. The relative atomic orbital projected band structure analysis in different nanosheets is consistent with the stochiometric composition of the surface QLs of respective materials. Typically, the Sb p-orbital contributions at the band edge of surface-capped layers can be viewed as a perturbation in atomic orbital interaction with respect to natural $Bi_2Se_3$. Therefore, presence of the $SbBiSe_3$ capping layer only marginally changes the CBM/VBM energies, and significantly larger changes in band edge energies can be observed in the presence of the $Sb_2Se_3$ capping layer.

Next, **Fig.3(d)** reveals that each surface Bloch states of natural $Bi_2Se_3$ (at Dirac point) is strongly localized at the corresponding surfaces, as depicted in **Fig.3(d)**. In contrast, the surface Bloch states are strongly delocalized away from the surface of $Sb_2Se_3/Bi_2Se_3$, and these Bloch states strongly hybridizes through bulk penetration, as shown in **Fig.4(d)**. **Fig.5(d) and Fig. 6(d)** suggest that for $SbBiSe_3/Bi_2Se_3$, irrespective of the configurations, the surface Bloch states are relatively more (less) surface localized compared to $Sb_2Se_3/Bi_2Se_3$ ($Bi_2Se_3$), with a marginal hybridization of surface Bloch states through bulk region. It should be noted that these observations are consistent with the surface QL contribution near the band edges of natural and surface-capped $Bi_2Se_3$.

The annihilation of the Dirac point of $Sb_2Se_3/Bi_2Se_3$ and $SbBiSe_3/Bi_2Se_3$, may be attributed to delocalization induced surface conducting state hybridization through the bulk region. The strong and moderate delocalization of surface Bloch states across the bulk allows notable interaction between the top and bottom surface states electronic wave functions of $Sb_2Se_3/Bi_2Se_3$ and $SbBiSe_3/Bi_2Se_3$, as depicted in **Fig.3(d)**, **Fig.4(d)**, **Fig.5(d)**, and **Fig.6(d).** This hybridization of surface states split the CBM and VBM and induce a finite bandgap at 'G' point [63], wherein the stronger surface state hybridization leads to larger energy bandgap opening in $Sb_2Se_3/Bi_2Se_3$.

Finally, the Fermi surface at 'G' point (at CBM) with spin projection profiles are considered for individual material systems and are depicted in **Fig.3(e)**, **Fig.4(e)**, **Fig.5(e)**, and **Fig.6(e)**. In general, for both natural and surface-capped $Bi_2Se_3$ nanosheets, the top and bottom Fermi surfaces are circular near the 'G' high-symmetry point of the BZ and tend to become hexagonal at higher energies. In this representation, the direction of the arrow represents the in-plane spin direction, wherein the Fermi surface represents the components of in-plane momenta ($k_x$ and $k_y$) at different energy iso-surfaces from the band edge, i.e. 'G' point. The in-plane spin projection profiles reveal that the top and bottom Fermi surfaces are characterized by perfect clockwise and anticlock wise spin-rotations, respectively in natural and $SbBiSe_3$ surface capped $Bi_2Se_3$. It should be noted that the calculated in-plane spin rotations in the top and bottom Fermi surfaces of natural $Bi_2Se_3$ in this work are consistent with previous theoretical report [53]. Thus, the in-plane spin direction reverses with the reversal of in-plane momentum (k) suggesting a spin-momentum locking for electrons in both top and bottom surface conducting states, implying completely spin-polarized unidirectional electronic transport in the top and bottom surfaces of these material systems. In contrast, no clear spin rotation features can be observed in $Sb_2Se_3/Bi_2Se_3$, suggesting existence of mixed spin electrons in the strongly delocalized surface electronic states in this case.

Thus, it can be surmised that for surface capped material systems, apart from the band edge energies, the perturbation introduced in the surface capping layers also tends to influence the surface localization of conducting states and their spin chirality.

### 3.3. Effects of Biaxial Strain on Electronic Properties of Natural and Surface Capped $Bi_2Se_3$ Nanosheets

In this sub-section, the effects of biaxial compressive (BC) and biaxial tensile (BT) strain on the surface electronic properties of natural and surface-capped $Bi_2Se_3$ slabs are considered, and the energy band structures under different applied BC and BT strains are presented as supplementary information (**Fig. S4 to Fig. S11**). In general, the applied biaxial strain significantly affects the CBM and VBM energies, wherein energy band structures gradually evolve with increasing strain, and a qualitatively similar nature of evolution can be observed with BC and BT strain for each material system. Typically, the applied BC strain (tensile stress) retains the gapless nature at the Dirac point. However, a notable modulation in valance band dispersion can be observed with increasing BC strain, wherein two secondary VBM near the 'G' high symmetry point (along 'G' to 'M' and 'G' to 'K' BZ paths) appears above the primary VBM at the 'G' point and severely compromises the linearity of E-k dispersion of VB around the 'G' point at higher BC strain. In contrast, the applied BT strain (compressive stress) retains the linearity of E-k dispersion near both CB and VB edges. However, increasing BT strain annihilates the Dirac point by inducing a finite energy bandgap (natural $Bi_2Se_3$ nanosheet) or enhances the existing energy bandgap (surface-capped $Bi_2Se_3$ nanosheets) at the 'G' point by shifting the CBM and VBM away from each other. At a higher BT strain, a secondary CBM appears along the 'G' to 'M' path of BZ, which remains sufficiently above the 'G' point throughout the BT strain range considered in this work. It should be noted that such finite surface energy bandgap opening at the 'G' point with tensile in-plane strain is consistent with the previous report [18].

Next, the comparative modulation in the energy band structure and DOS at large BC (5%) strain and BT (5%) strain with respect to the relaxed condition for natural and surface-capped $Bi_2Se_3$ nanosheets are presented in **Fig.7(a)**. **Fig.8(a) Fig. 9(a), Fig. 10(a),** and **Fig.7(b)**. **Fig.8(b) Fig. 9(b), Fig. 10(b),** respectively. Moreover, the effects of applied strain on these material systems are quantified from the band edge energy and bandgap variations with strain, as depicted in **Fig.7(c)**. **Fig.8(c) Fig. 9(c), Fig. 10(c),** and **Fig.7(d)**. **Fig.8(d) Fig. 9(d), Fig. 10(d),** respectively. Finally, to develop a physical insight about the effects of strain on atomistic properties, the ELF profiles of surface QL layers are considered at large BC (5%) strain and BT (5%) strain with respect to the relaxed condition are depicted in **Fig.7(e)**. **Fig.8(e) Fig. 9(e), Fig. 10(e)**.

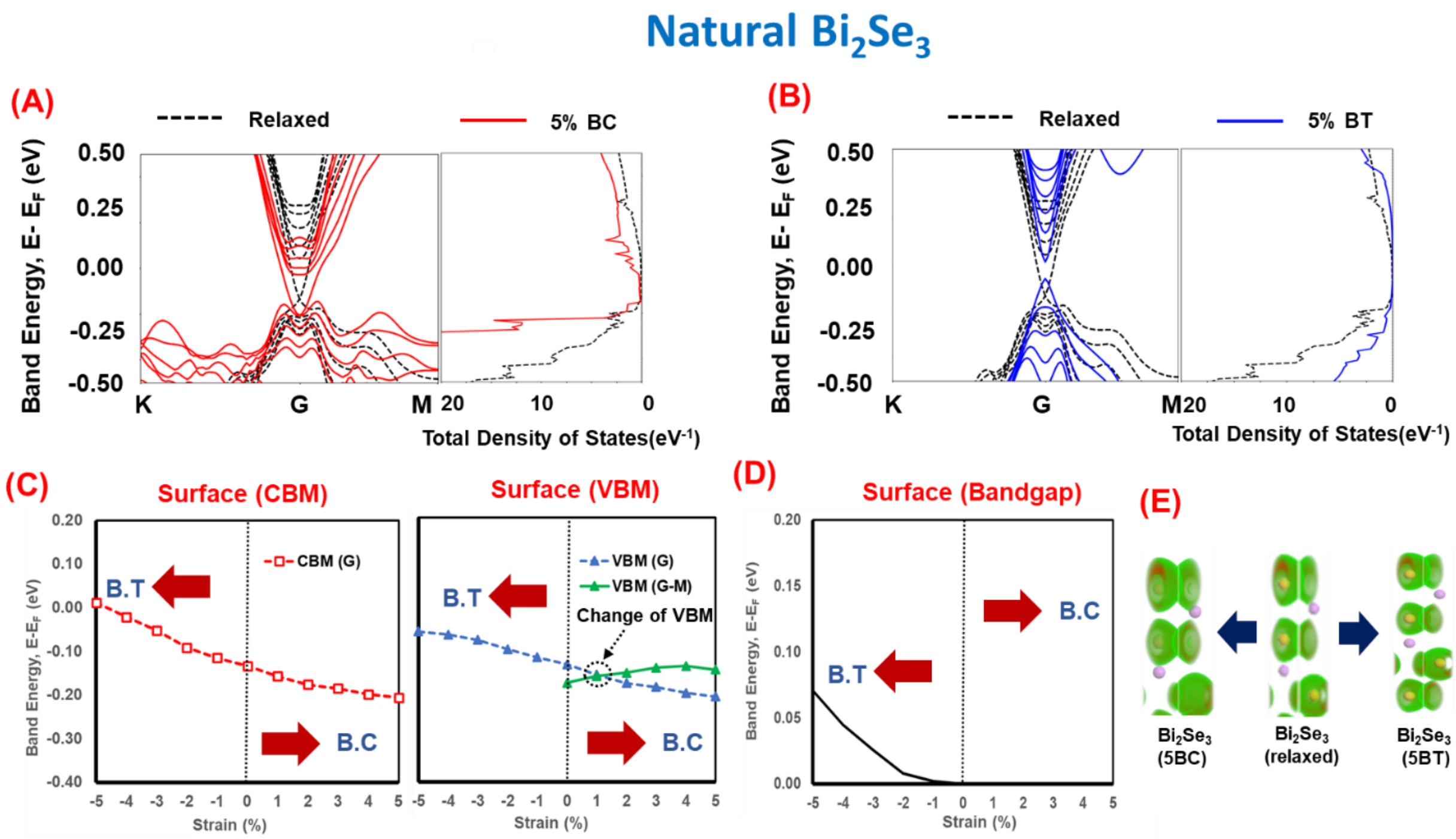

**Fig. 7.** Comparative plots of energy band structure and density of states of (a) relaxed and compressive strained, (b) relaxed and tensile strained natural $Bi_2Se_3$, plots of (c) band edge energy (d) bandgap variation of natural $Bi_2Se_3$ with applied strain, and comparative ELF profiles of (e) relaxed, compressive strained, and tensile strained natural $Bi_2Se_3$.

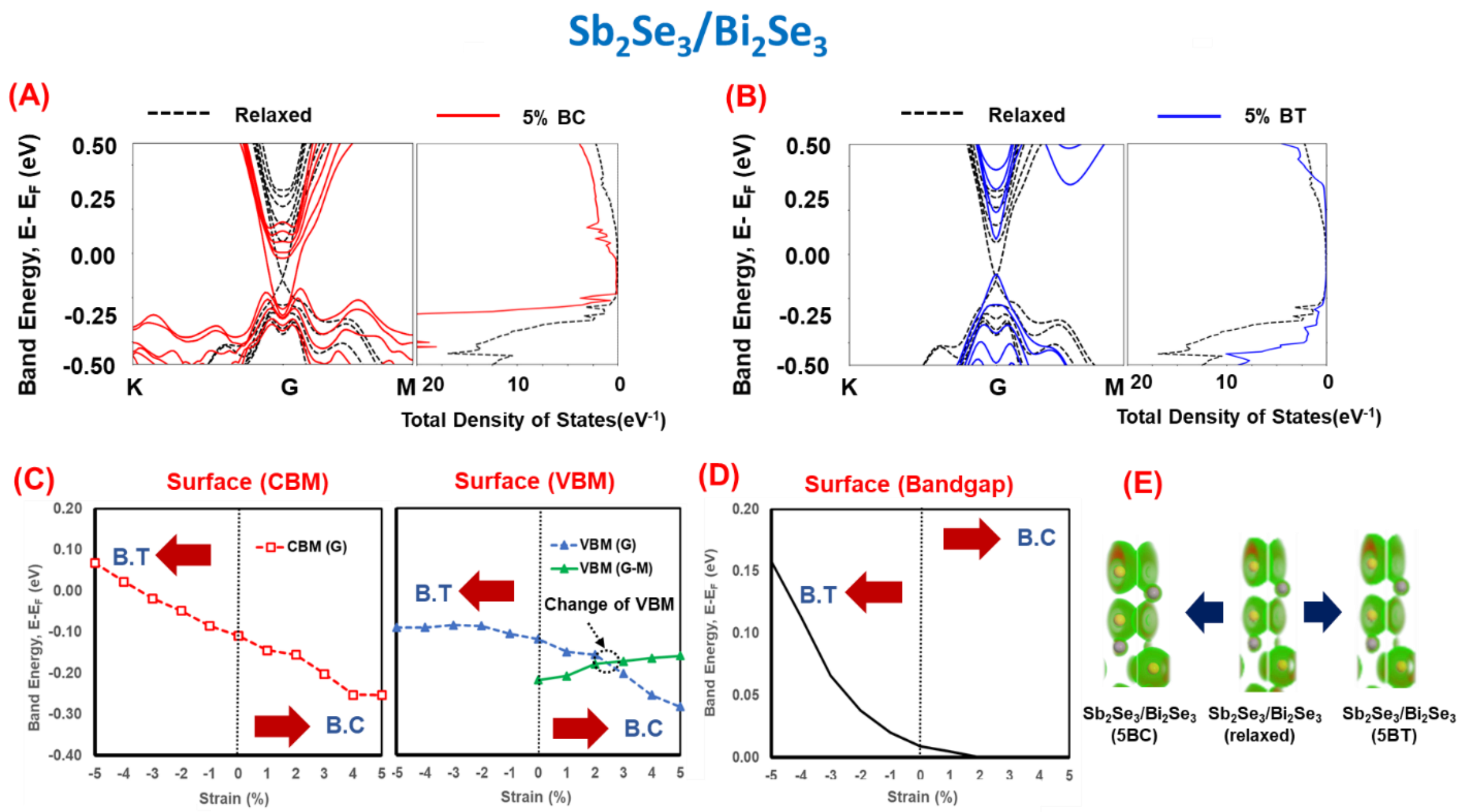

**Fig. 8.** Comparative plots of energy band structure and density of states of (a) relaxed and compressive strained, (b) relaxed and tensile strained $Sb_2Se_3/Bi_2Se_3$, plots of (c) band edge energy (d) bandgap variation of $Sb_2Se_3/Bi_2Se_3$ with applied strain, and comparative ELF profiles of (e) relaxed, compressive strained, and tensile strained natural $Sb_2Se_3/Bi_2Se_3$.

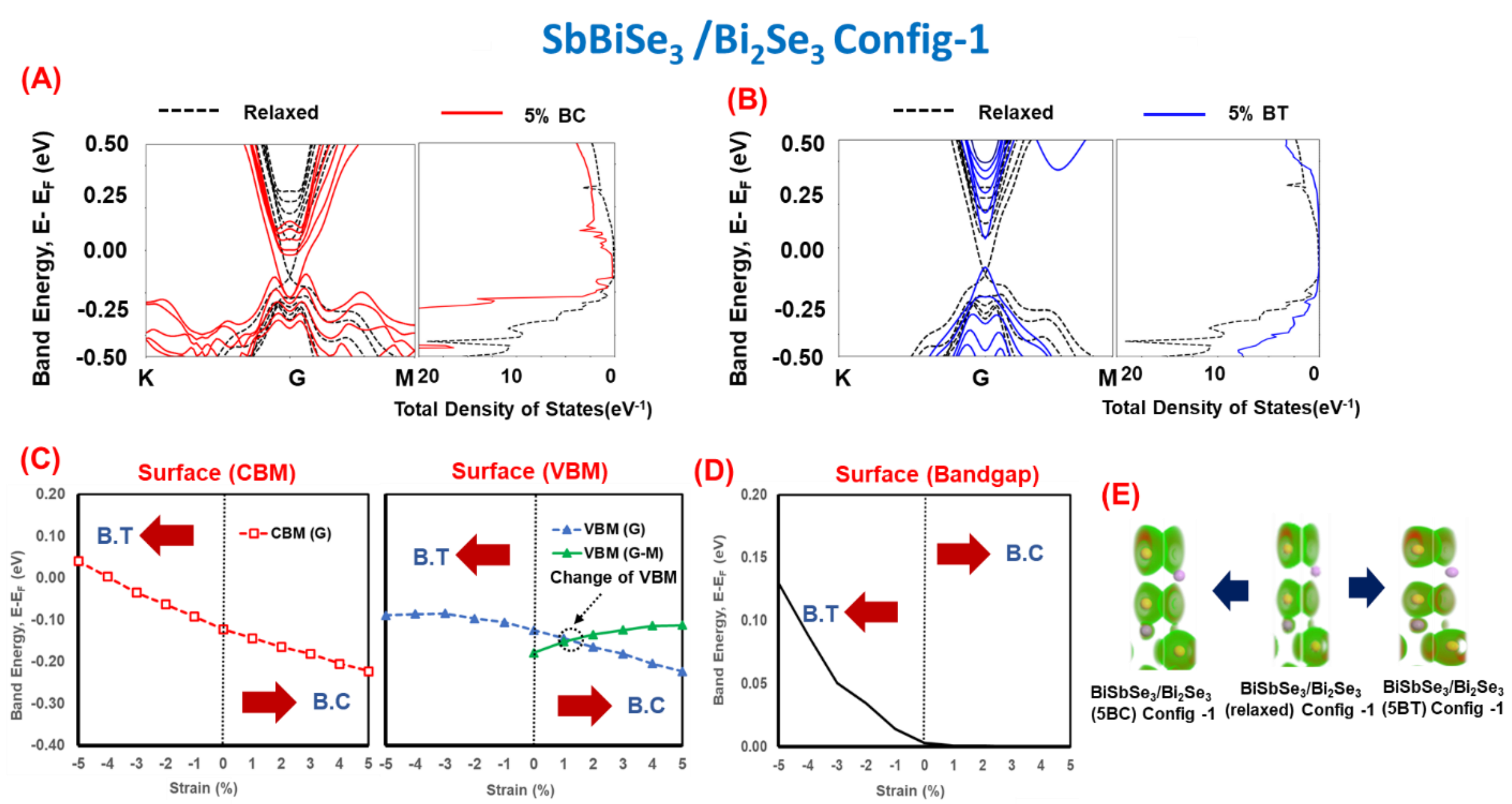

**Fig. 9.** Comparative plots of energy band structure and density of states of (a) relaxed and compressive strained, (b) relaxed and tensile strained $SbBiSe_3/Bi_2Se_3$ configuration-1, plots of (c) band edge energy (d) bandgap variation of $SbBiSe_3/Bi_2Se_3$ configuration-1 with applied strain, and comparative ELF profiles of (e) relaxed, compressive strained, and tensile strained natural $SbBiSe_3/Bi_2Se_3$ configuration-1.

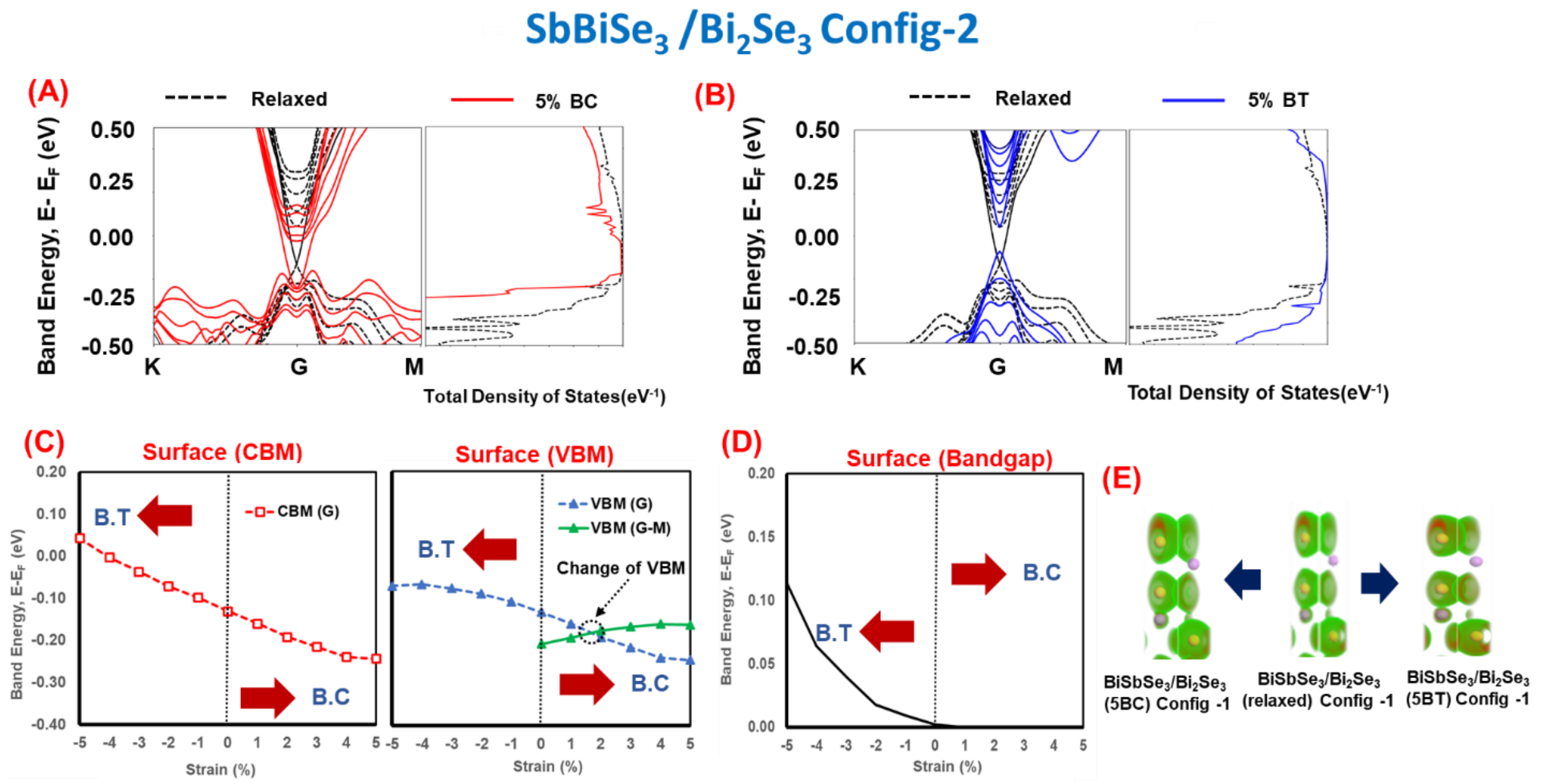

**Fig. 10.** Comparative plots of energy band structure and density of states of (a) relaxed and compressive strained, (b) relaxed and tensile strained $SbBiSe_3/Bi_2Se_3$ configuration-2, plots of (c) band edge energy (d) bandgap variation of $SbBiSe_3/Bi_2Se_3$ configuration-2 with applied strain, and comparative ELF profiles of (e) relaxed, compressive strained, and tensile strained natural $SbBiSe_3/Bi_2Se_3$ configuration-2.

In **Fig.7(a)**. **Fig.8(a) Fig. 9(a), Fig. 10(a),** the comparative analysis of energy band structures of natural and surface capped $Bi_2Se_3$ in relaxed condition and under large (5% BC and 5% BT) strain suggests a significant perturbation in the energy as well as E-k dispersion at 'G' point under applied large strain compared to relaxed condition. Specifically, conduction and valence band edges shift downwards (upward) in energy at large BC (BT) strain for individual material systems. This trend can be further corroborated by the fact that the CBM and VBM energies at 'G' point monotonically reduce from applied BT to BC strain, as illustrated in **Fig.7(c)**. **Fig.8(c) Fig. 9(c), Fig. 10(c)**. Throughout the range of applied BC and BT strain, the conduction band edge energy changes more sharply compared to valence band edge energy, which retain or restore the Dirac point at higher BC strain. It should be noted that such relative variation in the CBM and VBM energies of natural $Bi_2Se_3$ with strain is in good agreement with the previous theoretical reports [24].

In contrast to the primary VBM energy at 'G' point, the secondary VBM energy between 'G' to 'M' points increases with BC strain. This leads to the shift in VBM from 'G' point to between 'G' to 'M' points at higher BC strain that can be further verified from energy band structure evolution with BC strain (**Fig. S4, Fig.S6, Fig.S8, Fig. S10**).Consequently, for each material systems, an energy bandgap opening can be observed for BT strain, where the magnitude of bandgap opening is lowest (highest) for natural $Bi_2Se_3$ ($Sb_2Se_3/Bi_2Se_3$), as depicted in in **Fig.7(d)**.

**Fig.8(d) Fig. 9(d), Fig. 10(d)**. Moreover, an intermediate bandgap opening can be observed in $SbBiSe_3/Bi_2Se_3$, which is relatively higher in configuration-1. Thus, the relative magnitudes of BT strain induced bandgaps of different material systems are consistent with their relative bandgaps at the relaxed configuration, suggesting a uniform perturbation in atomic orbital interactions with applied strain in each case. On the other hand, the shift in band edge energies and change in curvature of E-k dispersion at band edges significantly influences the DOS profile leading to larger (smaller) DOS at the band edges with applied BC (BT) strain for individual material systems, as shown in **Fig.7(b)**. **Fig.8(b) Fig. 9(b), Fig. 10(b)**. Interestingly, owing to the loss of linearity of E-k dispersion in VB, a sharp increase in DOS can be observed at VB edge for large BC strain. **Fig.7(e)**. **Fig.8(e) Fig. 9(e), Fig. 10(e)** indicates that the application of BC and BT strain notably modulates the electronic distributions in surface QLs compared to their relaxed configuration. A careful observation reveals a relatively higher (lower) electron localization between the Se and Bi (Sb) atoms with applied BT (BC) strain, which is more prominent around the Sb atom. As discussed earlier, such perturbations in spatial electron distributions and, thereby, atomic orbital interactions in surface QL significantly influence the surface electronic properties.

Since, the application of biaxial strain modulates the atomic orbital interactions in surface QLs in natural and surface-capped $Bi_2Se_3$ nanosheet, it is expected to affect the surface localization of conducting states. Moreover, the previous reports have established that the spin-orbital coupling in $Bi_2Se_3$ is significantly influenced by the application of mechanical strain in the lattice [42-43], which suggests a strain-dependent spin chirality in these material systems. Subsequently, the evolution of out-of-plane projected surface Bloch states (at band edge) across the 6QL slabs, and Fermi-surfaces (at 'G' point) with its in-plane spin projections on momentum component with applied strain are considered, and are illustrated in **Fig.11** to **Fig.14**.

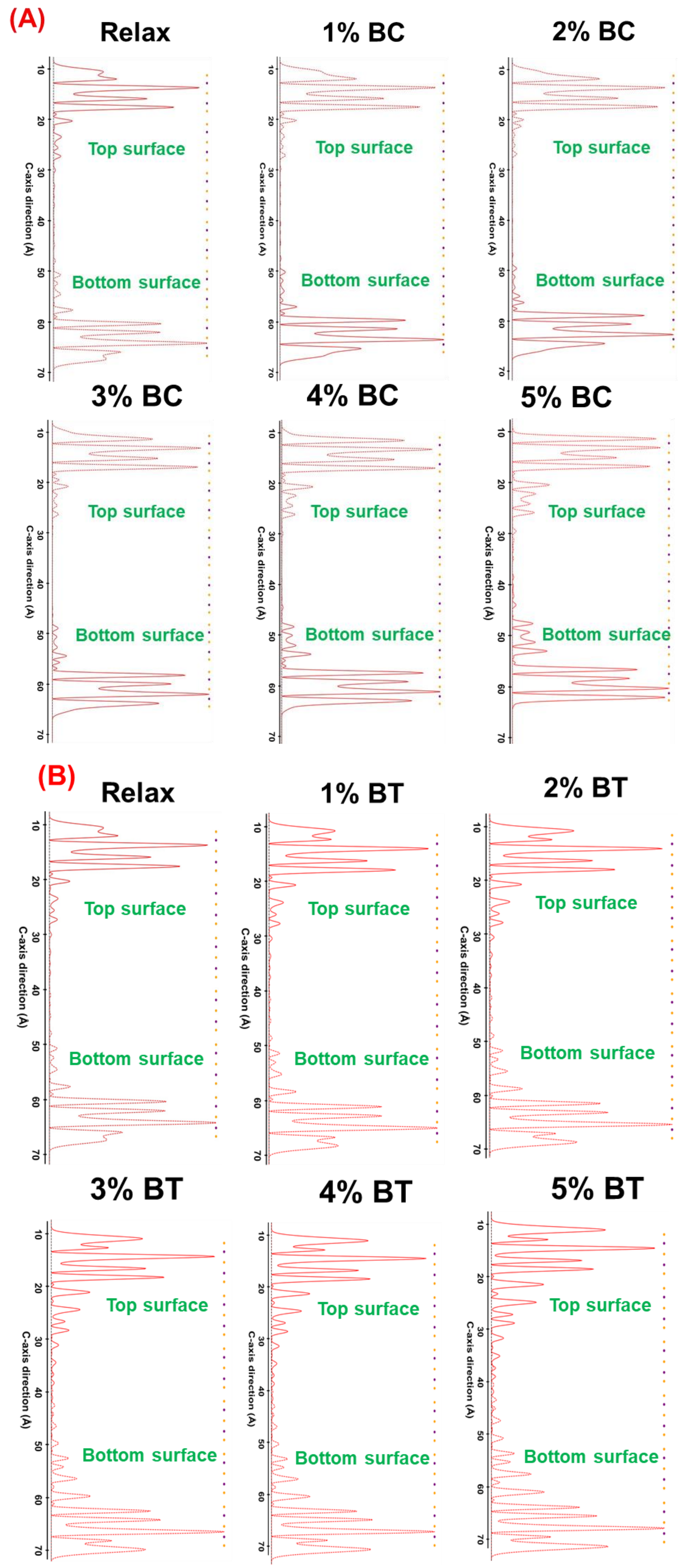
(A)
Relax
1% BC
2% BC
3% BC
4% BC
5% BC
(B)
Relax
1% BT
2% BT
3% BT
4% BT
5% BT
Top surface
Bottom surface
C-axis direction (A)
10
20
30
40
50
60
70

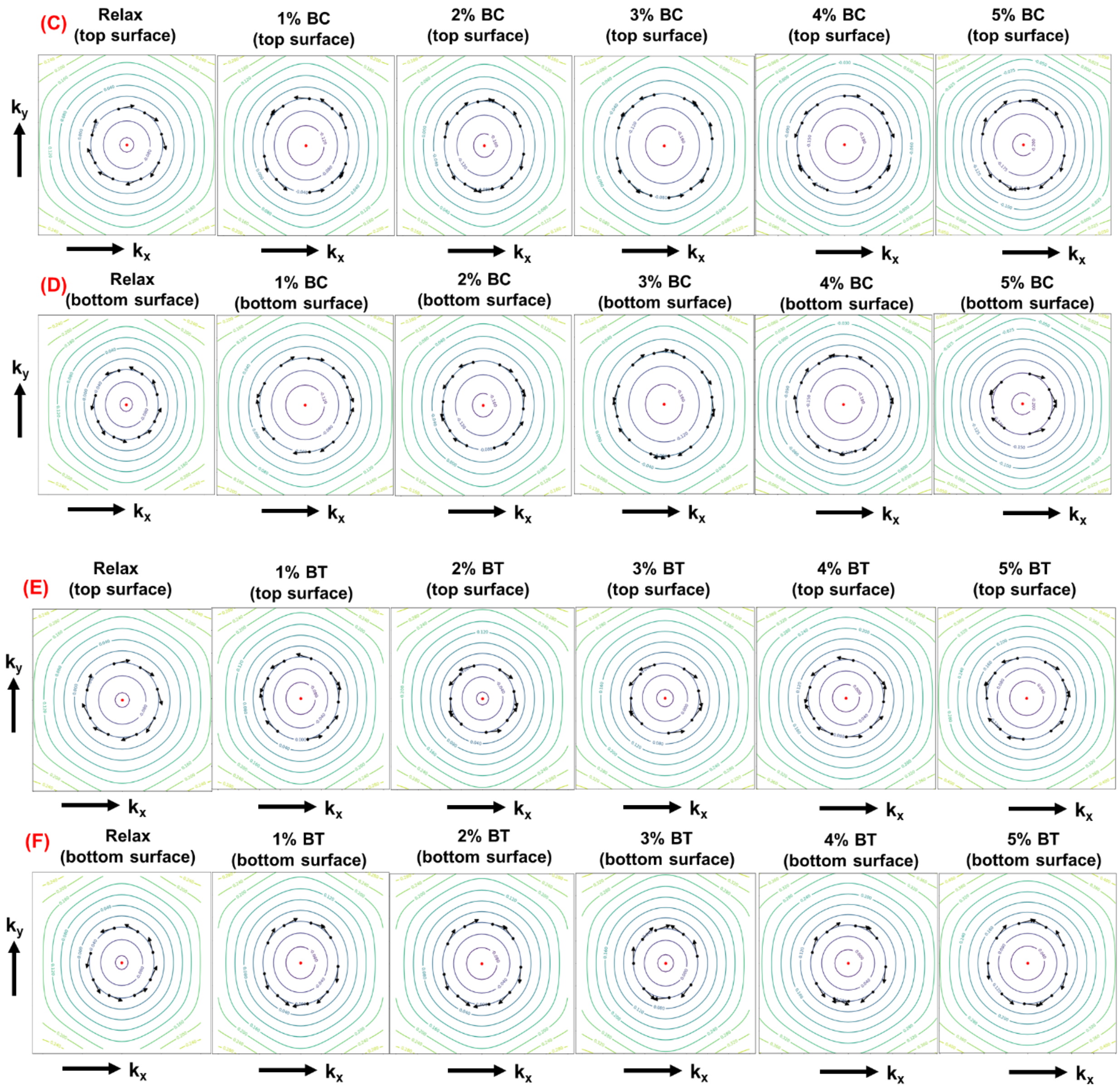

**Fig.11.** Plots of surface Bloch states projections (along C-axis) for applied (a) BC, (b) BT strain, Fermi surfaces on the Dirac cone (at G-point) for applied (c) BC strain at top surface, (d) BC strain at bottom surface, (e) BT strain at top surface, (d) BT strain at bottom surface of natural $Bi_2Se_3$.

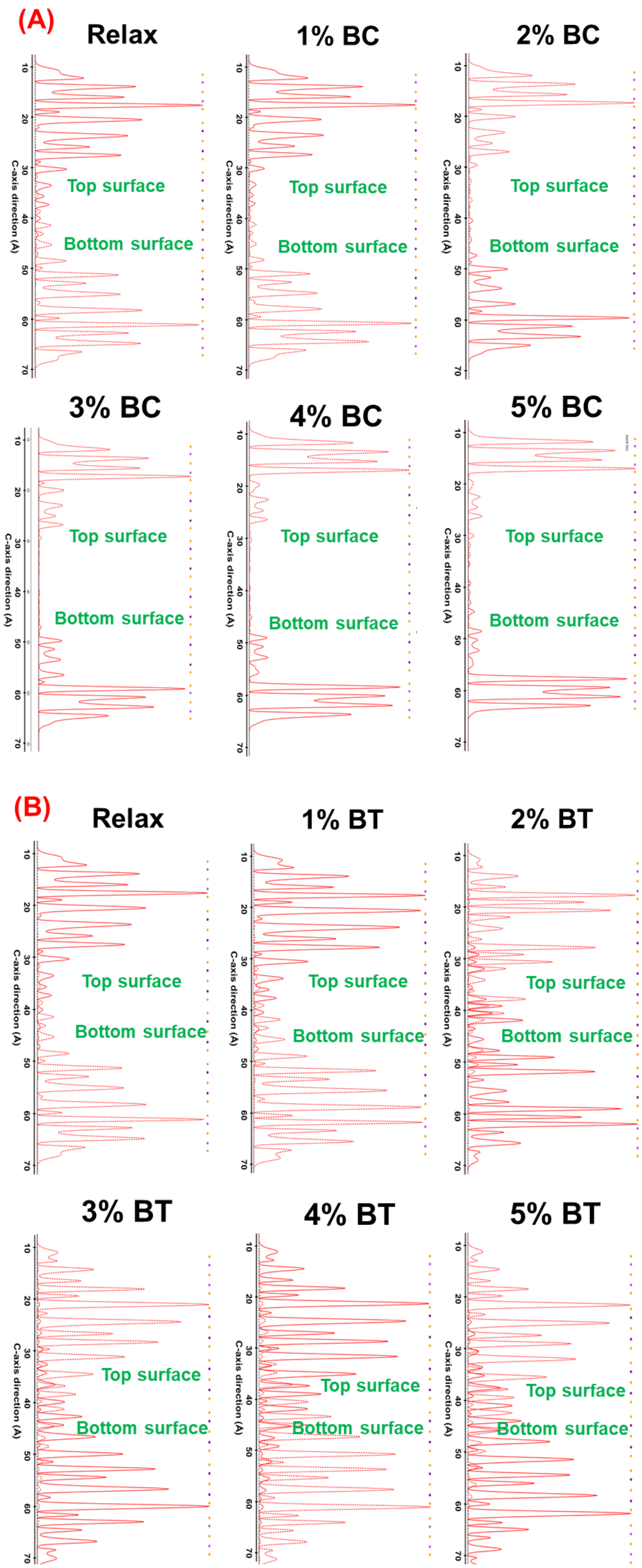
(A)
Relax
1% BC
2% BC
3% BC
4% BC
5% BC
(B)
Relax
1% BT
2% BT
3% BT
4% BT
5% BT
Top surface
Bottom surface
C-axis direction (A)
10
20
30
40
50
60
70

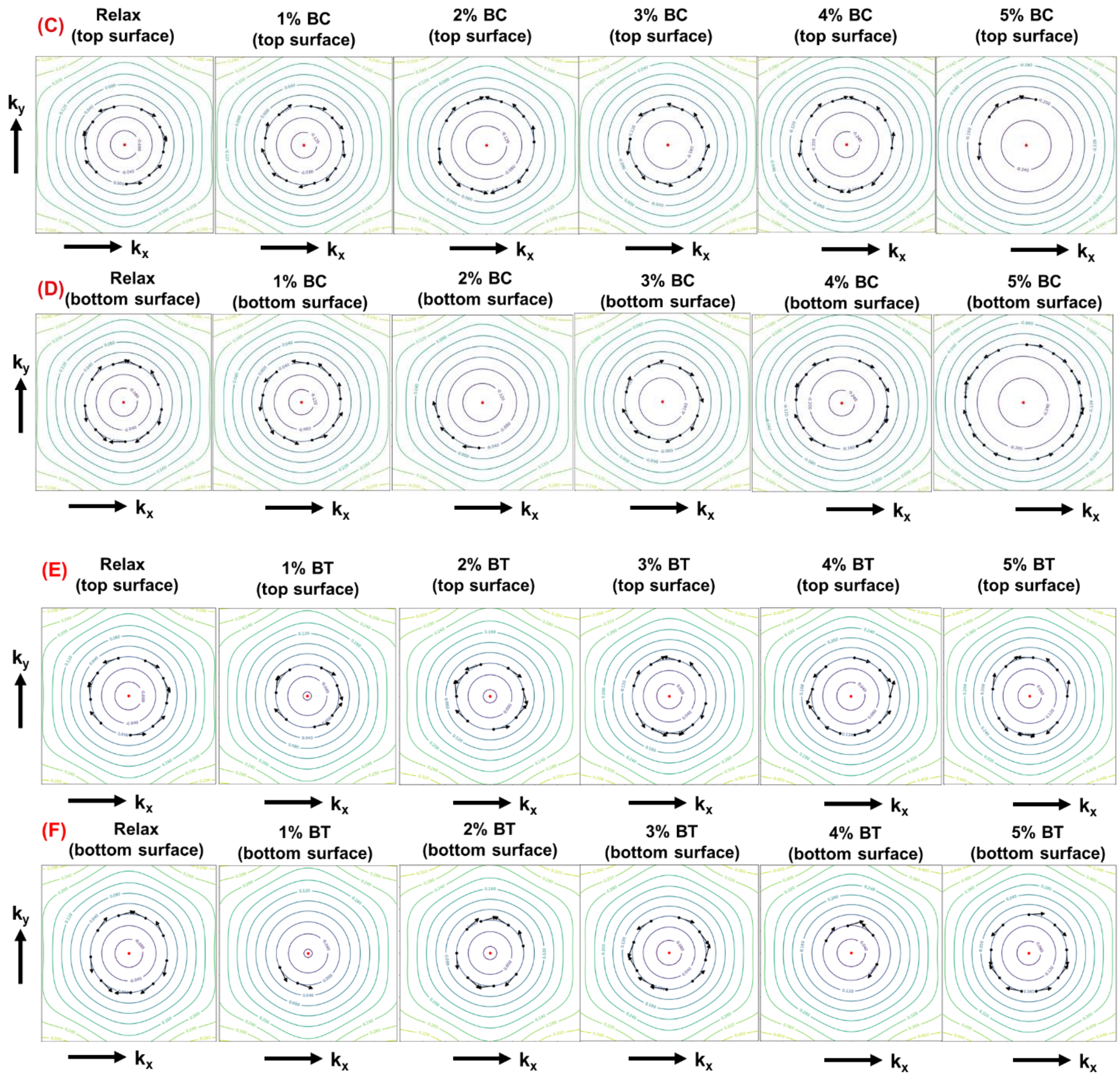

**Fig.12.** Plots of surface Bloch states projections (along C-axis) for applied (a) BC, (b) BT strain, Fermi surfaces on the Dirac cone (at G-point) for applied (c) BC strain at top surface, (d) BC strain at bottom surface, (e) BT strain at top surface, (d) BT strain at bottom surface of $Sb_2Se_3/Bi_2Se_3$.

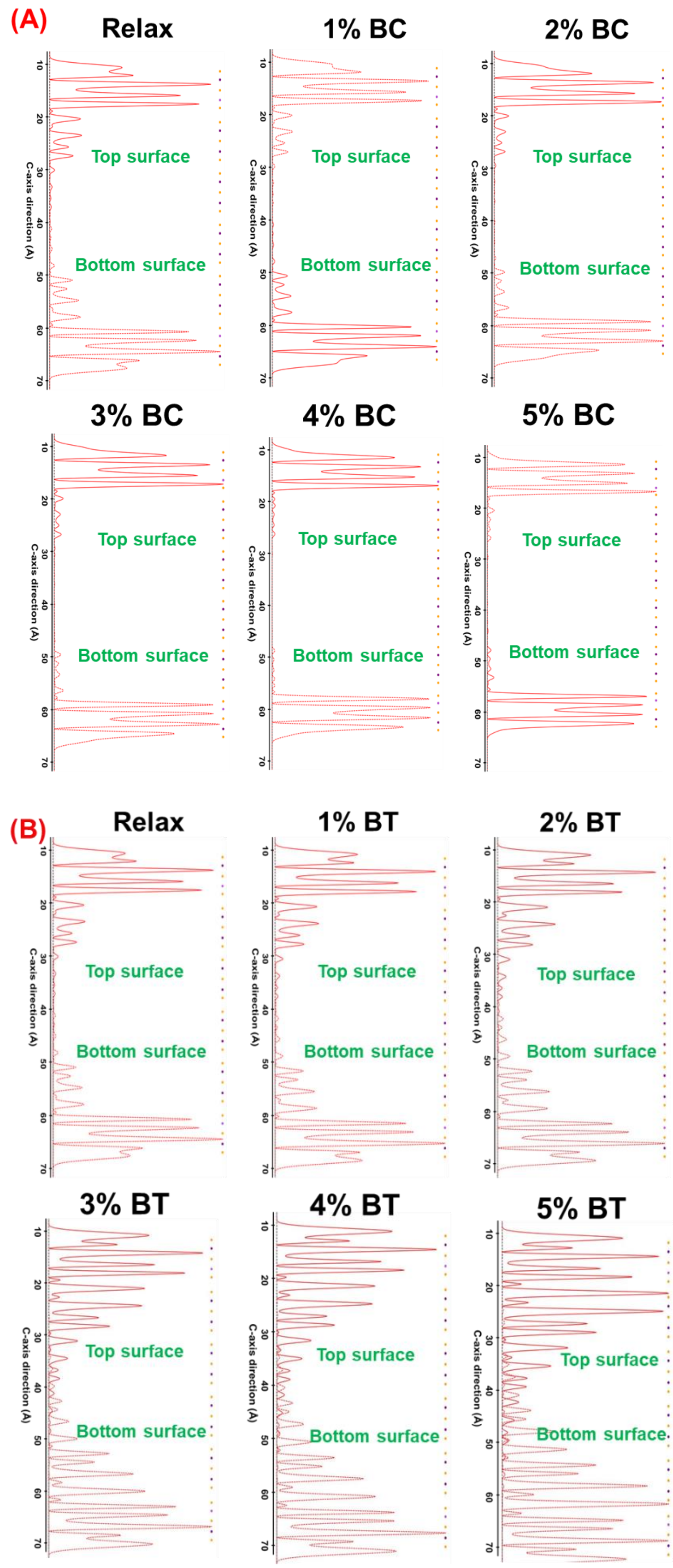

(A)
Relax
1% BC
2% BC
3% BC
4% BC
5% BC
Top surface
Bottom surface
C-axis direction (A)
(B)
Relax
1% BT
2% BT
3% BT
4% BT
5% BT
Top surface
Bottom surface
C-axis direction (A)

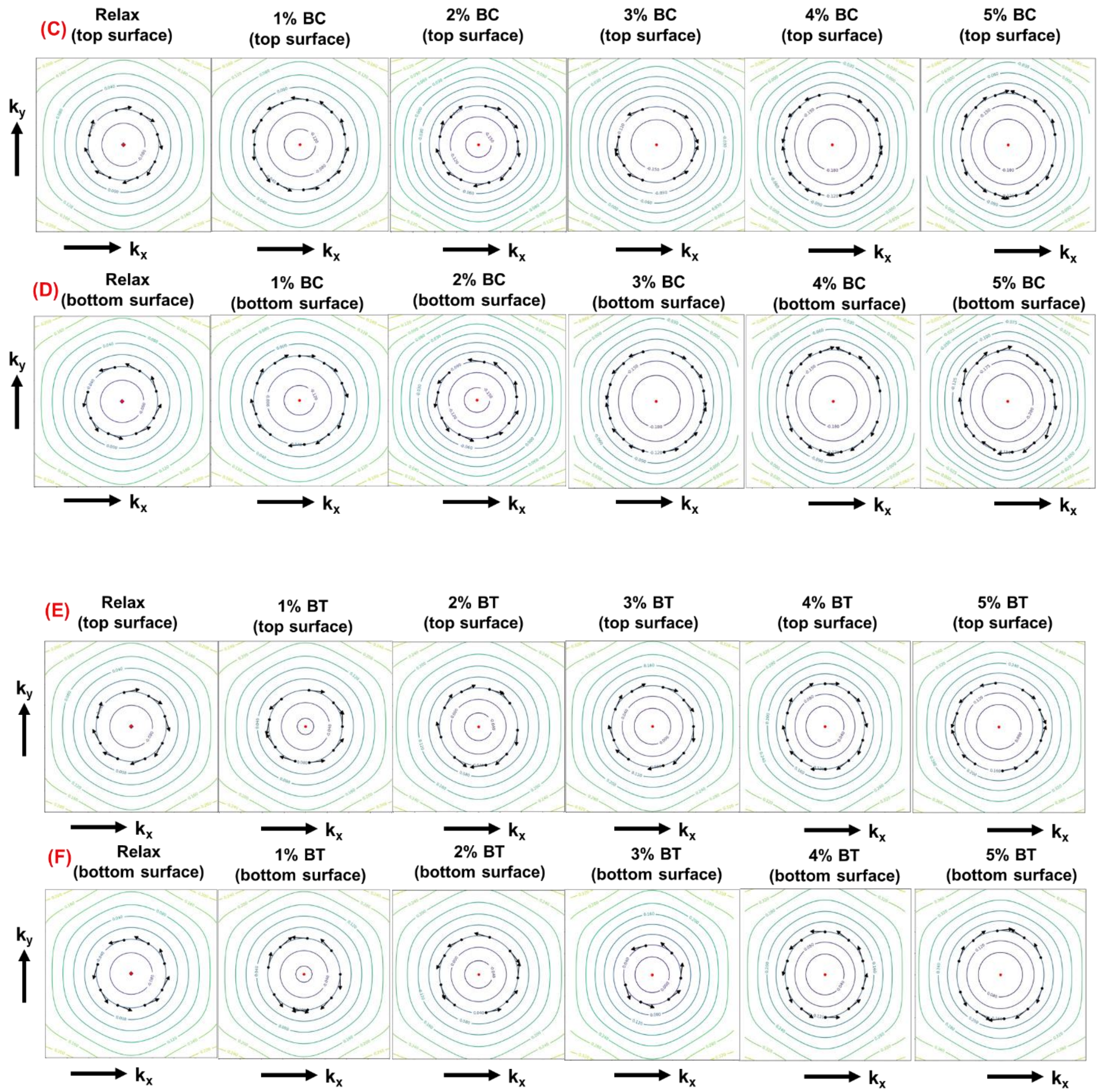

**Fig.13.** Plots of surface Bloch states projections (along C-axis) for applied (a) BC, (b) BT strain, Fermi surfaces on the Dirac cone (at G-point) for applied (c) BC strain at top surface, (d) BC strain at bottom surface, (e) BT strain at top surface, (d) BT strain at bottom surface of $SbBiSe_3/Bi_2Se_3$ configuration-1.

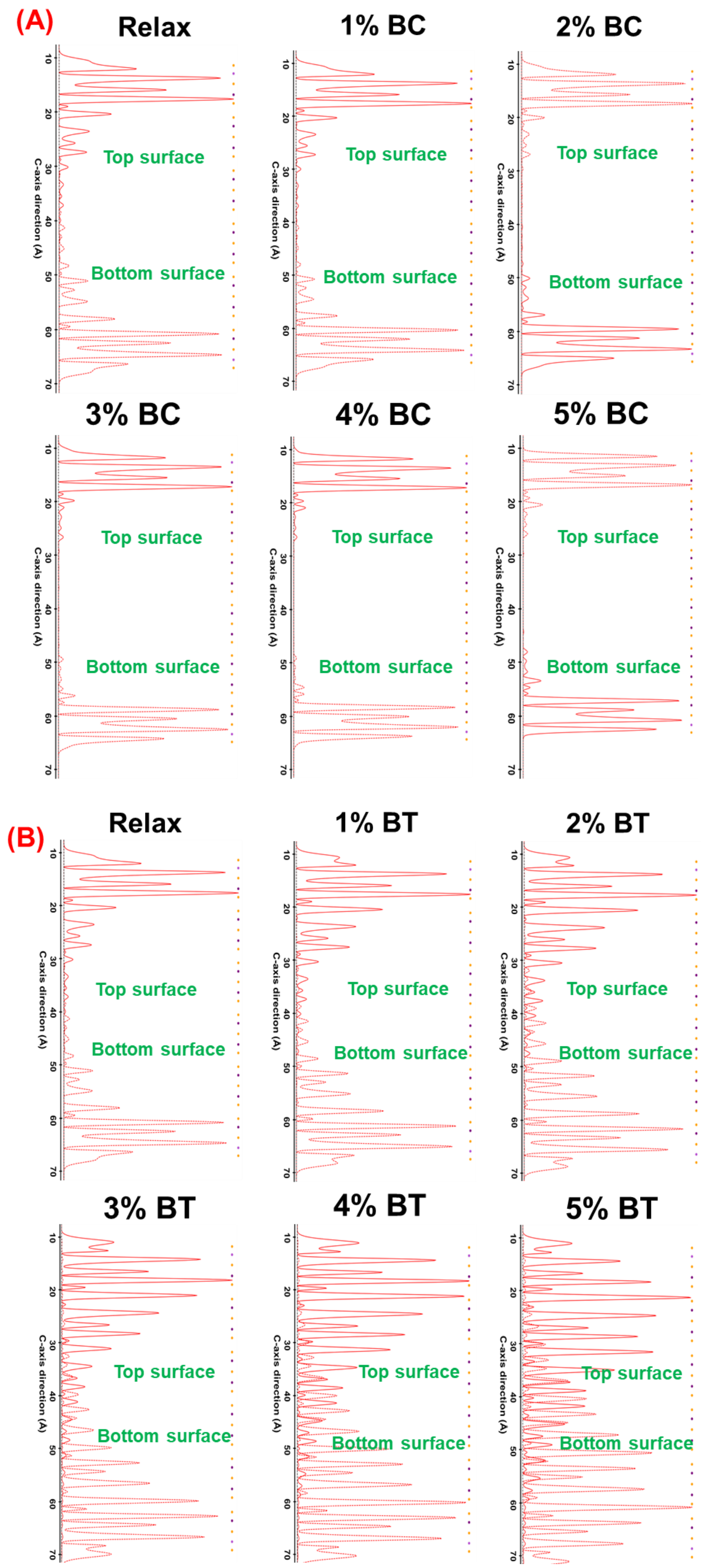
(A)
Relax
1% BC
2% BC
3% BC
4% BC
5% BC
Top surface
Bottom surface
C-axis direction (A)
10
20
30
40
50
60
70
(B)
Relax
1% BT
2% BT
3% BT
4% BT
5% BT

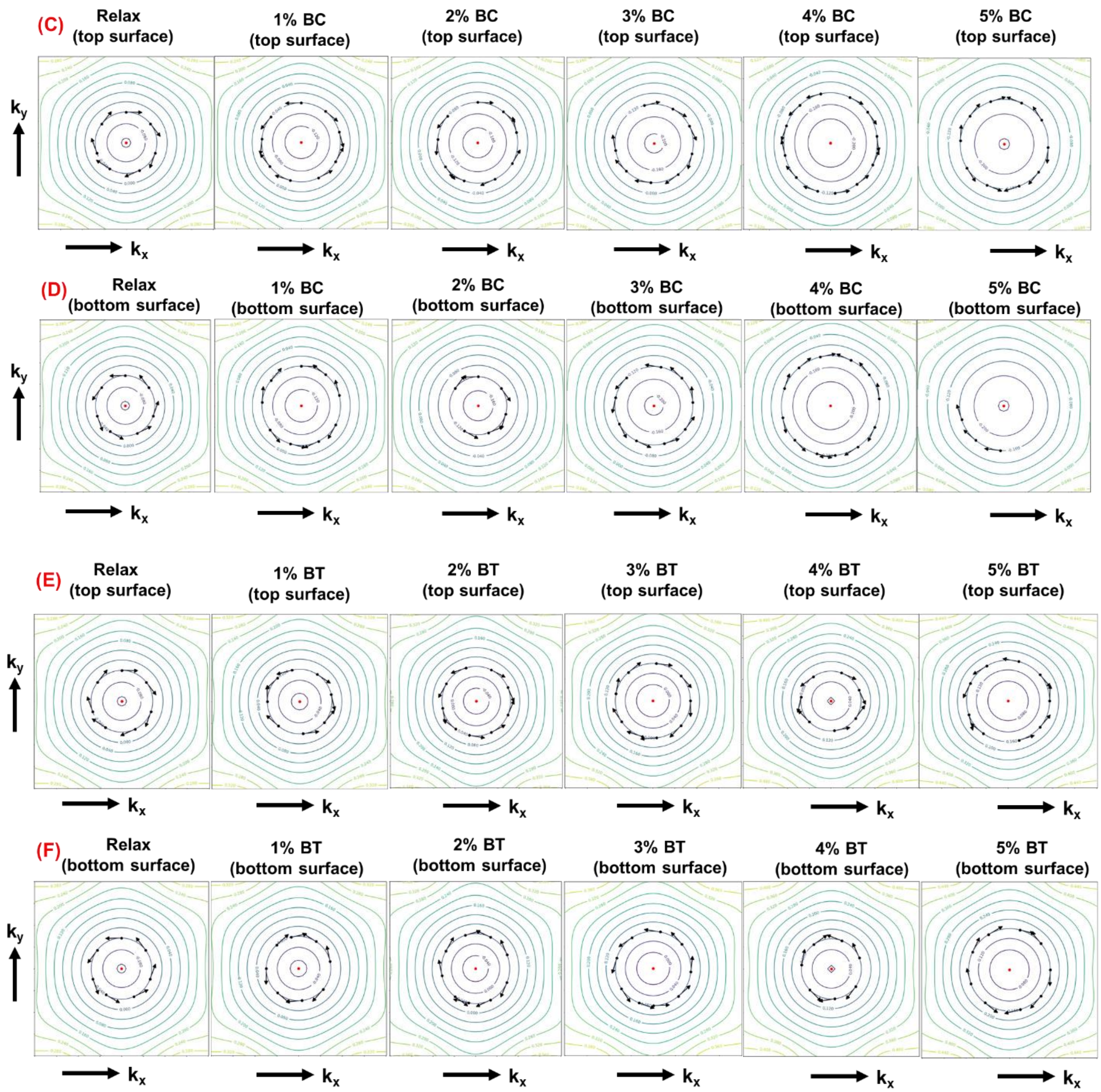

**Fig.14.** Plots of surface Bloch states projections (along C-axis) for applied (a) BC, (b) BT strain, Fermi surfaces on the Dirac cone (at G-point) for applied (c) BC strain at top surface, (d) BC strain at bottom surface, (e) BT strain at top surface, (d) BT strain at bottom surface of $SbBiSe_3/Bi_2Se_3$ configuration-2.

The results suggest that the applied strain significantly influences both surface localization of conducting states and their spin chirality. In general, the applied BT strain in any material system gradually delocalizes the surface conducting states, and eventually leads to strong bulk penetration for higher (>4%) BT strain, as depicted in **Fig.11(b), Fig.12(b), Fig.13(b), and Fig.14(b)**. This trend is most (least) prominent in $Sb_2Se_3/Bi_2Se_3$ (natural $Bi_2Se_3$), whereas an intermediate delocalization can be observed in $SbBiSe_3/Bi_2Se_3$. On the other hand, the BC strain has a distinct influence on the conducting surface state localization of individual material systems.

Specifically, **Fig.11(a)** indicates that in natural $Bi_2Se_3$, the applied BC strain effectively retains the localization of surface conducting states. Whereas, only a partial delocalization towards the bulk is observed at a higher (≥4%) BC strain. In $Sb_2Se_3/Bi_2Se_3$, the delocalized conducting surface states gradually localize at the surface with BC strain, and thereafter, a small delocalization can be observed at higher (~5%) BC strain, as shown in **Fig.12(a)**. However, **Fig.13(a) and Fig. 14(a)** indicates that in both the configurations of $Sb_2Se_3/Bi_2Se_3$, a conducting surface states gradually delocalizes with BC strain.

Therefore, in general, with the compression of the lattice (BT strain) the stronger atomic orbital interactions cause delocalization of surface states suggesting decaying insulating nature in the bulk of the material. This trend, along with the elimination of the surface Dirac point with in-plane lattice compression, can potentially diminish the distinct difference between the conductivities at the surface and the bulk of the materials in each case. In contrast, the presence of moderate expansion of the lattice (BC strain) weakens the strength of atomic orbital interactions, and subsequently, a clear surface state localization can be achieved even in materials with existing surface conducting state delocalization in relaxed condition. This trend along with the surface Dirac point restoration with lattice expansion can potentially enhances distinct difference between the surface and bulk conductivities in each case. As a result, the biaxial strain can be considered an effective means for tuning the relative difference in bulk and surface conduction in surface capped $Bi_2Se_3$, which is most effective for $SbBiSe_3$ surface capped $Bi_2Se_3$.

On the other hand, within the applied range of biaxial strain (5% BC to 5% BT), the top and bottom Fermi surfaces retain their circular energy iso-surface nature near the 'G' high-symmetry point of the BZ and transforms into hexagonal iso-surface nature away from the 'G' point in each material systems, which can be verified from **Fig.11(c)-(f), Fig.12(c)-(f), Fig.13(c)-(f), and Fig.14(c)-(f)**. However, a careful observation reveals that hexagonal nature becomes more (less) prominent, i.e. occurs at a lesser (higher) energy difference from the 'G' point with applied BC (BT) strain. This observation can be attributed to the fact that lattice expansion (BC strain) in real space, compresses the BZ in reciprocal space leading to early transition from circular to hexagonal features in Fermi surface, wherein an exact opposite trend can be anticipated for that lattice compression (BT strain) in real space.

In natural and surface-capped $Bi_2Se_3$ nanosheet, the Dirac point annihilation with applied BT strain can be correlated with the top and bottom surface conducting state interactions and sub sequent hybridization through bulk.

As the application of BT strain increases the delocalization of surface Bloch states that lead to hybridization of these states through bulk. Consequently, the increasing hybridization of delocalized surface Bloch states with BT strain induce a splitting between the band edges and open up energy bandgap at the ‘G’ point. In this context, the presence of delocalized surface Bloch states in surface-capped $Bi_2Se_3$ nanosheet further enhances their hybridizations with the BT strain, leading to larger bandgap opening in these materials, which is most prominent in $Sb_2Se_3$/ $Bi_2Se_3$. In contrast, the BC strain localizes the surface Bloch states and reduces their strength of interaction through bulk, which eventually eliminate surface conducting state localization and closes the energy bandgap of surface-capped $Bi_2Se_3$ nanosheet at a higher BC strain.

Next, with applied biaxial strain, the in-plane spin chirality is significantly influenced. In relaxed natural $Bi_2Se_3$, a precise clockwise and anticlock wise spin-rotations can be observed for top and bottom Fermi surfaces. However, the application of biaxial strain completely destroys this spin chirality leading to a mixed surface spin state at any applied BC and BT strain. In contrast, $Sb_2Se_3$/$Bi_2Se_3$ demonstrated a mixed surface spin-state in the relaxed condition, which remain in the similar spin state throughout the range of applied BT strain. However, the application of BC strain restores the spin-chirality (1%BC strain) and thereafter destroys (2%BC strain), and again reinstate the spin-chirality with reversed spin rotations in top and bottom surfaces (3% BC strain). For further increase in BC strain, the spin-chirality is lifted off and mixed surface spin-state is retained in $Sb_2Se_3$/$Bi_2Se_3$. The chiral top and bottom surface spin states of relaxed $SbBiSe_3$/$Bi_2Se_3$ in configuration-1 undergoes consecutive reversal in spin rotation (1% BC and 2% BC strains), and afterwards the spin chirality is destroyed with increasing BC strain. The chiral surface spin-state of relaxed $SbBiSe_3$/$Bi_2Se_3$ in configuration-1 is initially destroyed with applied BT strain, then get restored (3% BT and 4% BT strain), and thereafter again destroyed. In contrast, the chiral surface spin-state of relaxed $SbBiSe_3$/$Bi_2Se_3$ in configuration-2 is initially destroyed, then restored (3% BC and 3% BT strain), and again destroyed with applied BC and BT strain. Therefore, application of suitable biaxial strain in surface capped $Bi_2Se_3$ can be consider an efficient strategy for tuning the spin-momentum locking and thereby spin-polarization of electrons in the surface state.

## 4. Conclusion

This work presents a comprehensive analysis of the effects of biaxial strain on the surface electronic properties in the natural, $Sb_2Se_3$ surface capped, and $SbBiSe_3$ surface capped $Bi_2Se_3$ nanosheets. The results indicate that the introduction of surface capping layers as well as application of biaxial strain has significant influence on the top and bottom surface Bloch state hybridization through bulk. The interplay between these two factors can induce emergent effects on the surface electronic properties of these material systems. The introduction of surface capping layer hybridizes the surface conducting state by delocalization, which distort the linearity of E-k dispersion and simultaneously annihilates the Dirac point by energy bandgap opening. The application of biaxial compressive strain closes down the surface energy bandgap and restores surface localization of conducting states in surface capped $Bi_2Se_3$, whereas an exact opposite trend is observed under biaxial tensile strain. The most prominent surface conducting state hybridization in presence of $Sb_2Se_3$ surface capping ensure largest bandgap opening in relaxed and biaxially tensile strained condition. The spin chirality and spin-momentum locking are preserved after introduction of $SbBiSe_3$ surface capping, wherein the same are completely destroyed in presence of $Sb_2Se_3$ surface capping in $Bi_2Se_3$. Interestingly, the results demonstrate that the spin-momentum locking can be restored and the in-plane spin rotation can also be reversed by applying suitable biaxial strain in surface capped $Bi_2Se_3$, wherein no such feature can be observed in natural $Bi_2Se_3$. However, the spin chirality and spin-momentum locking in $SbBiSe_3$ surface-capping is more responsive to the applied strain.

In essence, the comparative study suggests that the surface electronic properties and spin-texture of $SbBiSe_3$ surface-capped $Bi_2Se_3$ can be efficiently tuned by applying biaxial strain, which could be promising for novel electronic, optoelectronic, and spintronic applications. Finally, the study offers important theoretical insights into the surface capping and strain co-engineering in $Bi_2Se_3$ with respect to their surface electronic property modulation, which can potentially guide the experimental realization of tailor-made electronic and spintronic properties in $Bi_2Se_3$ in future for specific technological applications.

**Acknowledgements**

The computational research for this work was carried out in the Computational Nanomaterials and Nanodevices Research Lab (CNNRL) of the EEE Department, BITS-Pilani, Hyderabad Campus. Banasree Sadhukhan acknowledges Department of Science and Technology, Government of India, for financial support with reference no DST/WISE-PDF/PM-4/2023 under WISE Post-Doctoral Fellowship programme to carry out this work.